\documentclass[acmtog,anonymous=false,review=false,authorversion=false]{acmart}
\acmSubmissionID{369}

\pdfoutput=1

\usepackage{booktabs}
\usepackage{colortbl}
\usepackage{graphicx}
\usepackage{subcaption}
\usepackage{changepage}
\usepackage{wrapfig}
\usepackage{enumitem}
\usepackage{mathtools} 
\usepackage{siunitx} 

\usepackage{dblfloatfix}    

\newcommand{\edit}[1]{{#1}}

\newcommand{\editTwo}[1]{{#1}}

\usepackage{algorithm}
\usepackage[noend]{algpseudocode}

\algnewcommand{\LineComment}[1]{\Statex $\quad\;\;$\(\triangleright\) #1}

\definecolor{WHITE}{gray}{1.0}
\definecolor{GRAY}{gray}{0.9}
\definecolor{MATBLUE}{rgb}{0,0.4470,0.7410}
\definecolor{MATRED}{rgb}{0.8500,0.3250,0.0980}
\definecolor{MATYELLOW}{rgb}{0.9290,0.6940,0.1250}
\definecolor{MATPURPLE}{rgb}{0.4940,0.1840,0.5560}
\definecolor{MATGREEN}{rgb}{0.4660,0.6740,0.1880}

\newcommand{\eg}{\textit{e.g.,}~}
\newcommand{\ie}{\textit{i.e.,}~}

\pdfpageattr {/Group << /S /Transparency /I true /CS /DeviceRGB>>}

\let\orgautoref\autoref

\def\secnospace~{\S{}}
\renewcommand{\autoref}
        {\def\equationautorefname{Eq.}%
         \def\figureautorefname{Fig.}%
         \def\subfigureautorefname{Fig.}%
         \def\algorithmautorefname{Alg.\@}%
         \def\Itemautorefname{Item}%
         \def\tableautorefname{Table}%
         \def\sectionautorefname{\secnospace}%
         \def\subsectionautorefname{\secnospace}%
         \def\subsubsectionautorefname{\secnospace}%
         \def\chapterautorefname{\secnospace}%
         \def\partautorefname{Part}%
         \orgautoref}

\def\BibTeX{{\rm B\kern-.05em{\sc i\kern-.025em b}\kern-.08em
    T\kern-.1667em\lower.7ex\hbox{E}\kern-.125emX}}

\let\SS\undefined 
\let\vv\undefined 

\newcommand{\EE}{{\bf E}}
\newcommand{\FF}{{\bf F}}

\newcommand{\II}{{\bf I}}
\newcommand{\JJ}{{\bf J}}

\newcommand{\LL}{{\bf L}}
\newcommand{\MM}{{\bf M}}

\newcommand{\RR}{{\bf R}}
\newcommand{\SS}{{\bf S}}

\newcommand{\XX}{{\bf X}}

\newcommand{\ff}{{\bf f}}

\newcommand{\qq}{{\bf q}}

\newcommand{\vv}{{\bf v}}
\newcommand{\xx}{{\bf x}}

\newcommand{\zz}{{\bf z}}

\newcommand{\dispdot}[2][.05ex]{\dot{\raisebox{0pt}[\dimexpr\height+#1][\depth]{$#2$}}}

\newcommand{\dq}{\dot\qq}
\newcommand{\dqm}{\dq_m}
\newcommand{\dqr}{\dq_r}
\newcommand{\ddq}{\ddot\qq}

\newcommand{\ddqr}{\ddq_r}

\newcommand{\Jmr}{\JJ_{mr}}
\newcommand{\Jar}{\JJ_{{\alpha}r}}
\newcommand{\Jam}{\JJ_{{\alpha}m}}
\newcommand{\Jsx}{\JJ_{sx}}
\newcommand{\Jaz}{\JJ_{{\alpha}z}}
\newcommand{\Jzm}{\JJ_{zm}}
\newcommand{\Jxm}{\JJ_{xm}}
\newcommand{\Jax}{\JJ_{{\alpha}x}}

\newcommand{\JaO}[1]{^S\JJ_{{\alpha_{#1}}\text{o}}^\text{NN}}
\newcommand{\JaI}[1]{^S\JJ_{{\alpha_{#1}}\text{i}}^\text{NN}}
\newcommand{\dJaO}[1]{^S\dispdot\JJ_{{\alpha_{#1}}\text{o}}^\text{NN}}
\newcommand{\dJaI}[1]{^S\dispdot\JJ_{{\alpha_{#1}}\text{i}}^\text{NN}}

\newcommand{\Jbase}{\JJ_\text{base}}
\newcommand{\Jori}{\JJ_\text{ori}}
\newcommand{\Jins}{\JJ_\text{ins}}
\newcommand{\dJbase}{\dispdot\JJ_\text{base}}
\newcommand{\dJori}{\dispdot\JJ_\text{ori}}
\newcommand{\dJins}{\dispdot\JJ_\text{ins}}

\newcommand{\dJmr}{\dispdot{\JJ}_{mr}}
\newcommand{\dJar}{\dispdot{\JJ}_{{\alpha}r}}
\newcommand{\dJam}{\dispdot{\JJ}_{{\alpha}m}}
\newcommand{\dJsx}{\dispdot{\JJ}_{sx}}
\newcommand{\dJaz}{\dispdot{\JJ}_{{\alpha}z}}
\newcommand{\dJzm}{\dispdot{\JJ}_{zm}}
\newcommand{\dJxm}{\dispdot{\JJ}_{xm}}

\newcommand{\Ma}{\MM_\alpha}

\newcommand{\Mm}{\MM_m}
\newcommand{\fa}{\ff_\alpha}

\newcommand{\fm}{\ff_m}

\newcommand{\xO}{\xx_\text{ori}}
\newcommand{\xI}{\xx_\text{ins}}
\newcommand{\dxO}{\dot{\xx}_\text{ori}}
\newcommand{\dxI}{\dot{\xx}_\text{ins}}
\newcommand{\dxa}{\dot\xx_\alpha}
\newcommand{\ddxa}{\ddot\xx_\alpha}

\newcommand{\xas}[1]{^S\!\xx_{\alpha_{#1}}}

\newcommand{\xOs}{^S\!\xO}
\newcommand{\xIs}{^S\!\xI}

\newcommand{\sO}{s_\text{ori}}
\newcommand{\sI}{s_\text{ins}}

\newcommand{\dxOw}{^W\!\dxO}
\newcommand{\dxIw}{^W\!\dxI}
\newcommand{\dxaw}{^W\!\dxa}

\newcommand{\vBw}{^W\!\vv_\text{base}}
\newcommand{\vOw}{^W\!\vv_\text{ori}}
\newcommand{\vIw}{^W\!\vv_\text{ins}}

\newcommand{\vOwrel}{^W\!\!\vv_\text{ori}^\text{rel}}

\newcommand{\vOsrel}{^S\!\vv_\text{ori}^\text{rel}}
\newcommand{\vIsrel}{^S\!\vv_\text{ins}^\text{rel}}

\newcommand{\W}{W\!} 
\newcommand{\SO}[2]{\prescript{#1}{#2}\RR}
\newcommand{\SE}[2]{\prescript{#1}{#2}\EE}

\newcommand{\Zero}{\mathbf{0}}

\newcommand{\rsize}[1]{\mathbb{R}^{#1}}
\newcommand{\Rsize}[2]{\mathbb{R}^{#1 \times #2}}

\newcommand{\teaser}{
\begin{teaserfigure}
  \centering
  \subcaptionbox{\label{fig:teaserRun}}{
    \includegraphics[height=1.4in,trim={21.5in 5in 15in 4in},clip]{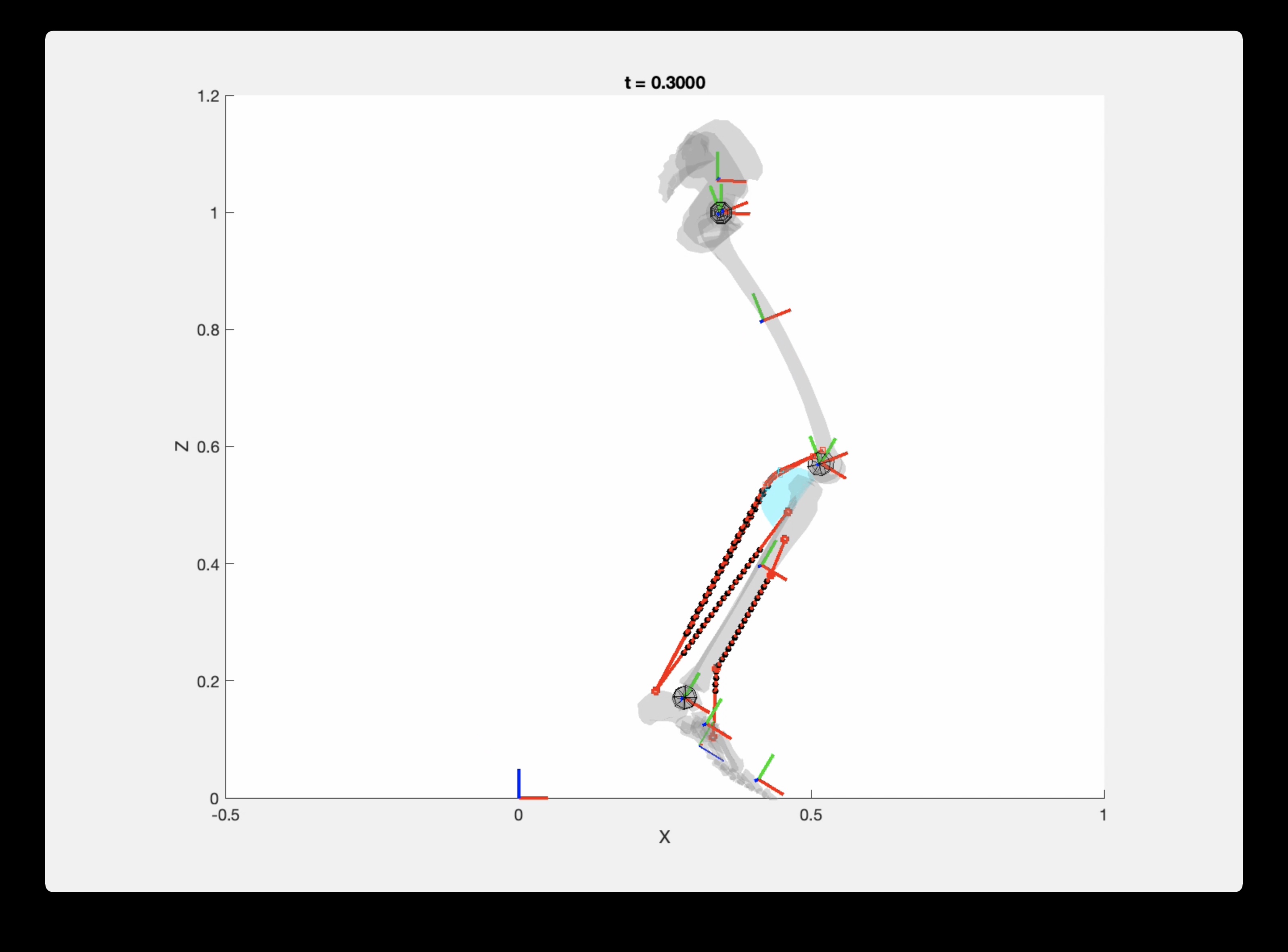}
  }
  \subcaptionbox{\label{fig:teaserFlick}}{
    \includegraphics[height=1.0in,trim={0in 0in 8in 0in},clip]{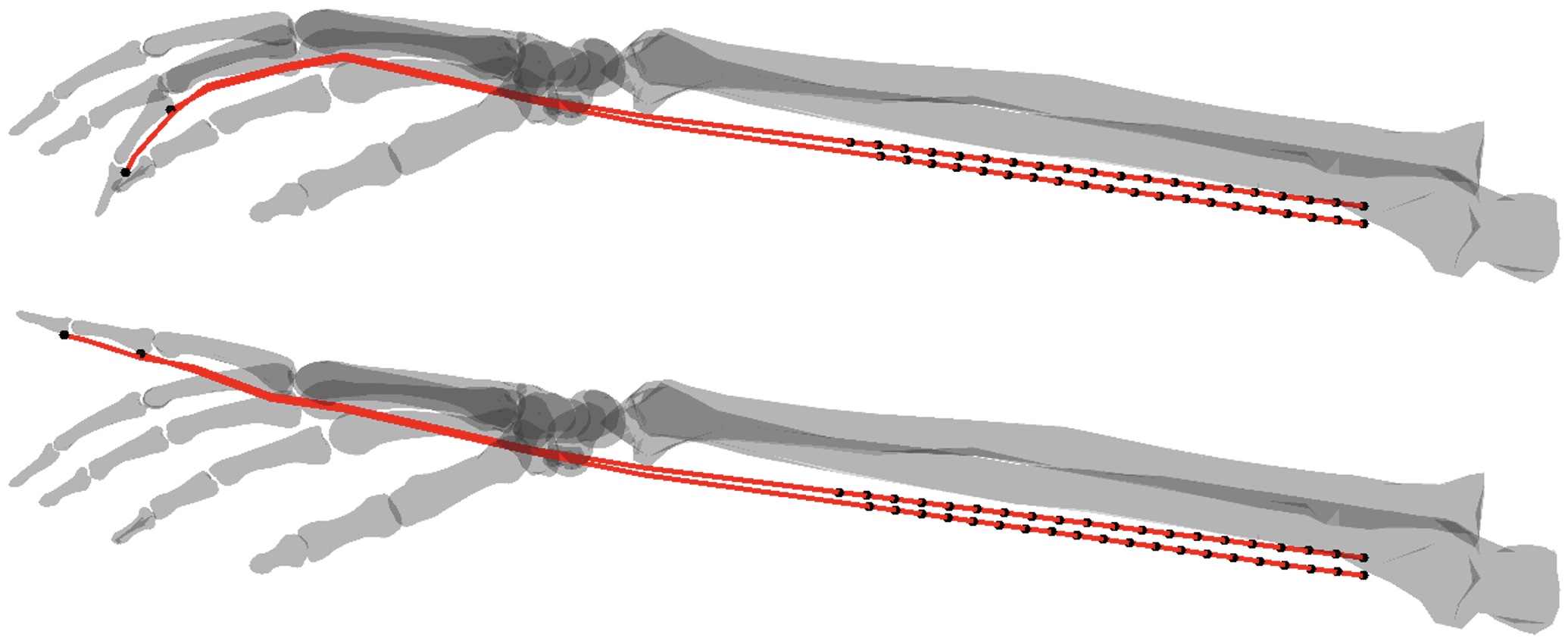}
    \vspace{0.2in}
  }
  \subcaptionbox{\label{fig:teaserHill}}{
  	\includegraphics[height=1.4in,trim={2.6in 6.4in 16.2in 1.9in},clip]{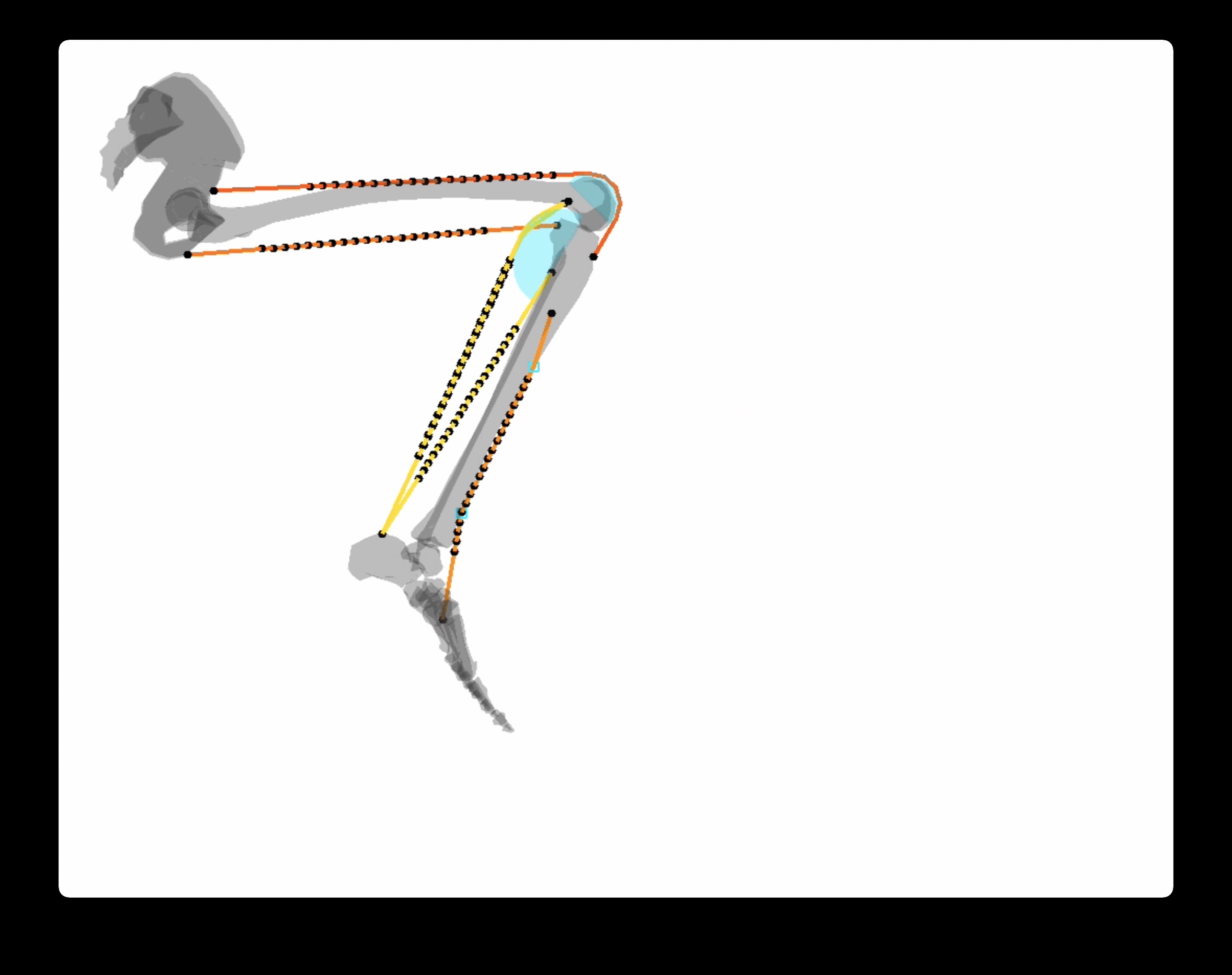}
  }
  \subcaptionbox{\label{fig:teaserReach}}{
   	\includegraphics[height=1.4in,trim={0in 0in 0in 0in},clip]{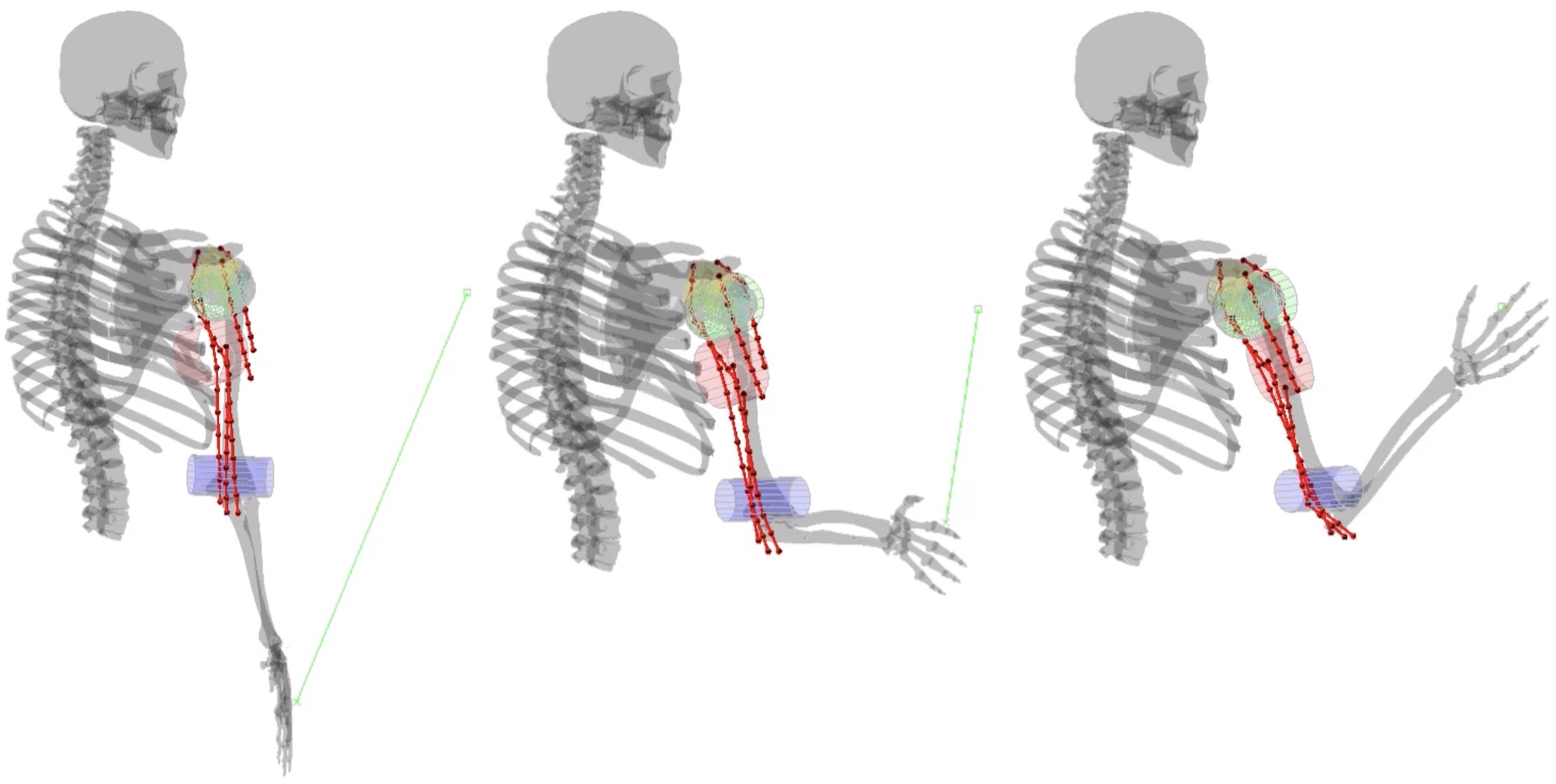}
  }
  \caption{
  \edit{
  Muscle inertia
  (a) changes the inverse dynamics result of running motion by up to 40\%, and 
  (b) stabilizes the simulation.
  Our framework
  (c) handles Hill-type muscles, complex joints, and higher-order integration, and
  (d) works flawlessly with the adjoint method for computing the simulation derivatives.
  }
  }
\end{teaserfigure}
}

\newcommand{\figConcrete}{
  \begin{figure}[tb]
    \centering
    \includegraphics[width=0.7\columnwidth,trim={0in 0in 0in 0in},clip]{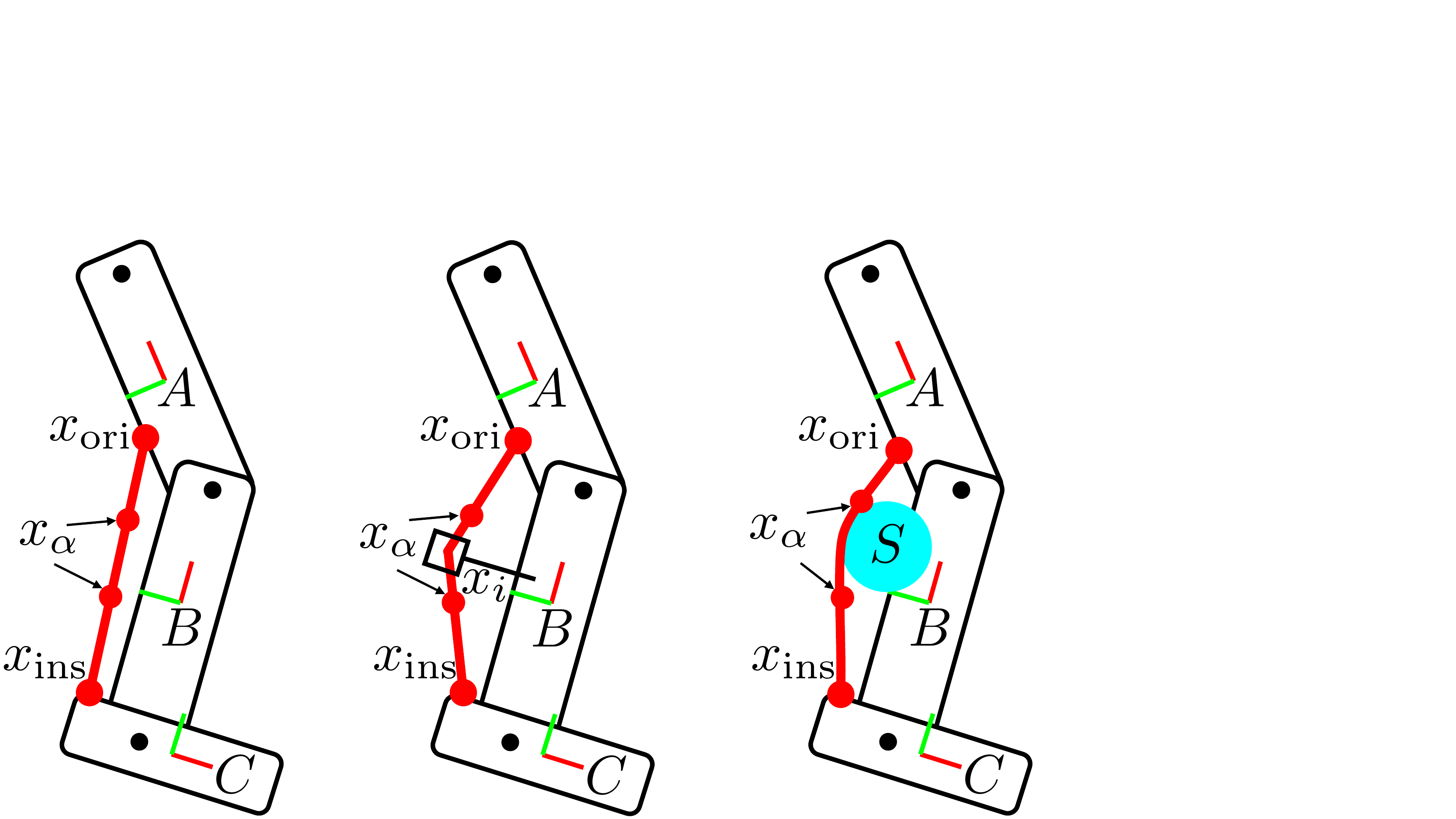}
    \caption{
      Concrete running example for Types I, II, and III muscles.
      In all cases, there are three bones and one muscle.
      The origin is on body $A$, and the insertion is on body $C$.
      Type II muscle has a path point on body $B$, and Type III muscle has a wrapping surface $S$ defined with respect to body $B$.
    }
    \label{fig:concrete}
  \end{figure}
}

\newcommand{\figWrapCollision}{
  \begin{figure*}[b!]
    \centering
    \subcaptionbox{Detaching \& attaching muscle\label{fig:wrapAnim}} {
    	\includegraphics[height=1.0in,trim={0in 0in 0in 0in},clip]{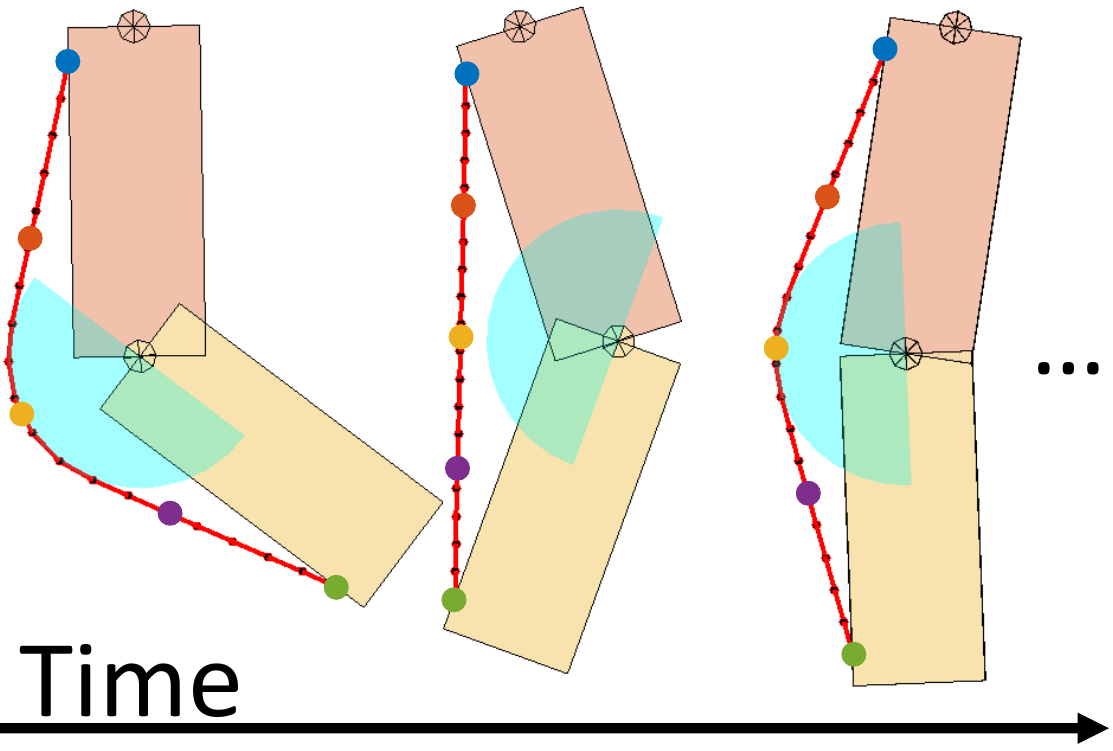}
    }
    \hspace{0.15in}
    \subcaptionbox{\editTwo{With library}\label{fig:wrapEnergyExisting}} {
    	\includegraphics[height=1.1in,trim={0.3in 0in 1.0in 0.8in},clip]{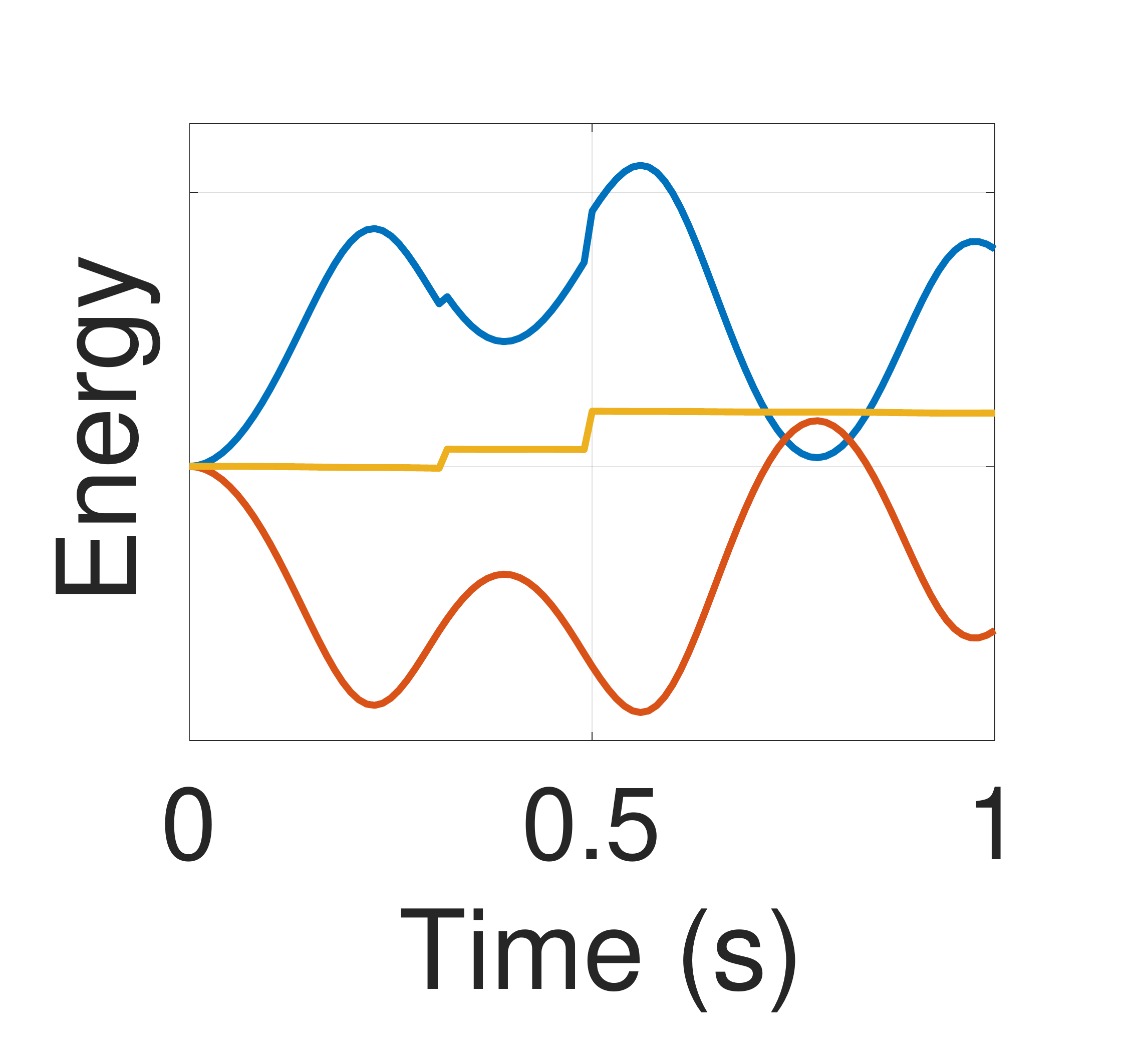}
    }
    \subcaptionbox{\editTwo{Ours}\label{fig:wrapEnergyOurs}} {
    	\includegraphics[height=1.1in,trim={0.3in 0in 1.0in 0.8in},clip]{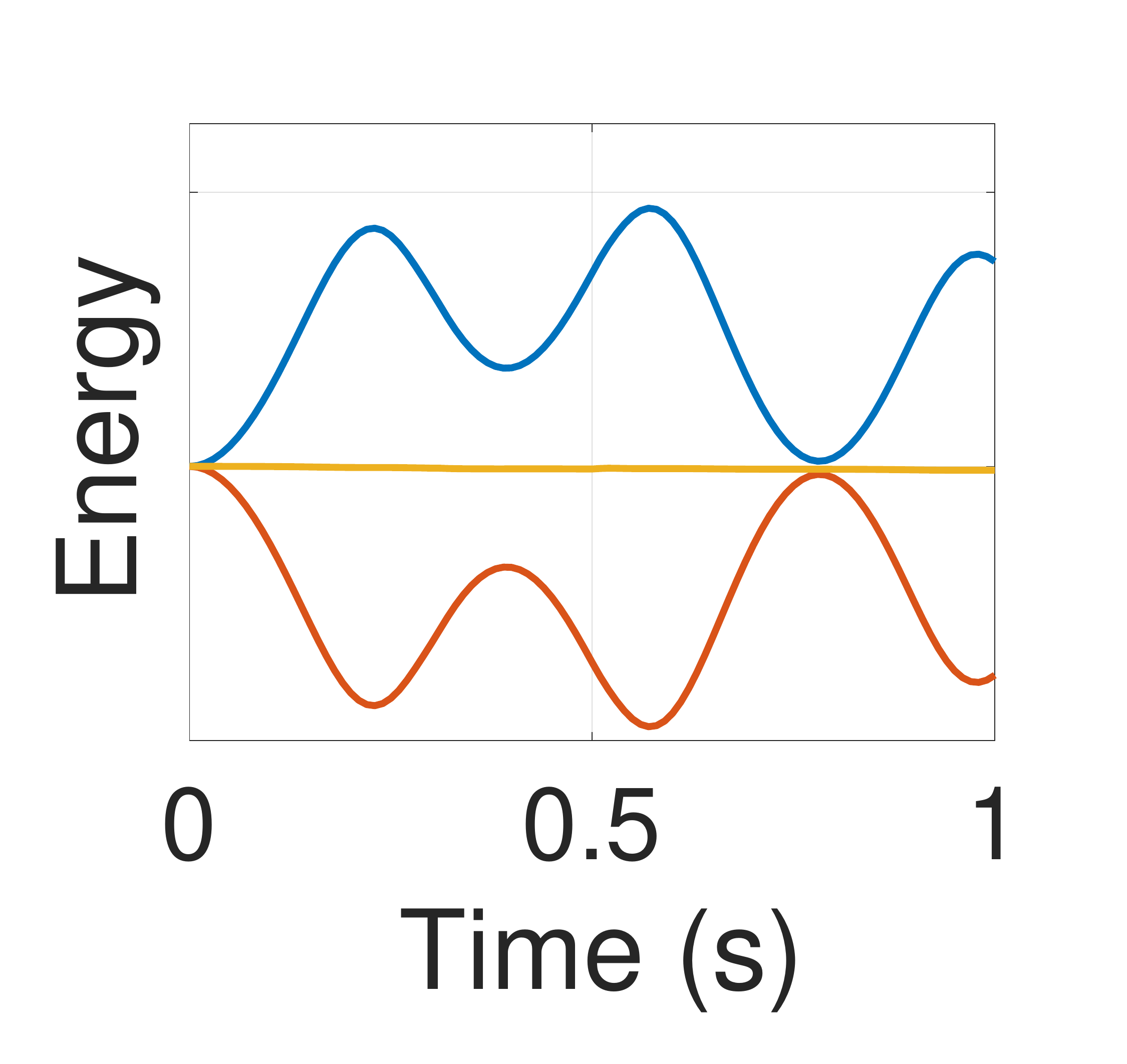}
    }
    \hspace{0.15in}
    \subcaptionbox{\editTwo{Positions}\label{fig:wrapX}} {
    	\vspace{0.5em}
    	\includegraphics[height=0.96in,trim={0.3in 0.2in 0.9in 0.8in},clip]{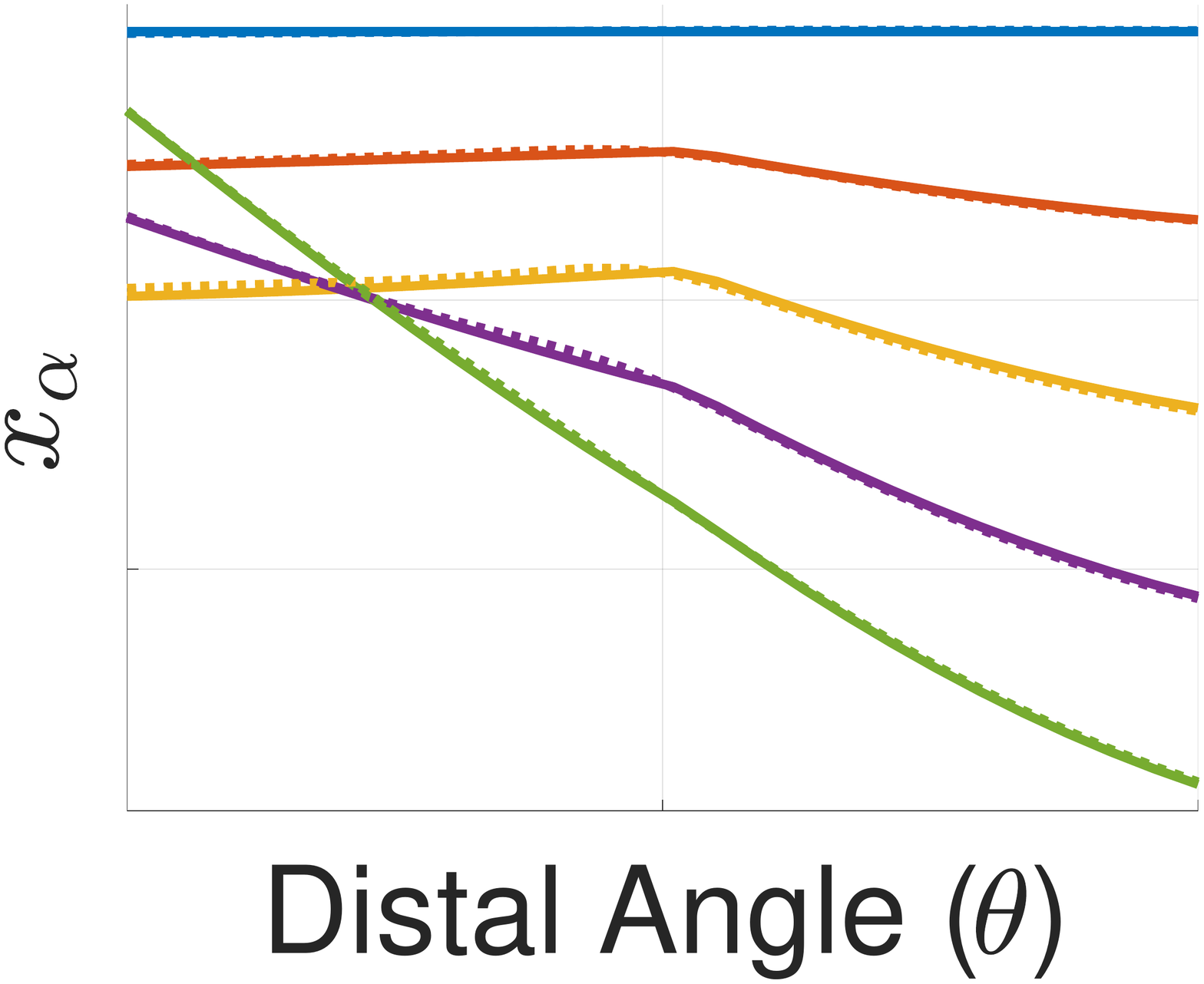}
    }
    \subcaptionbox{\editTwo{Derivatives}\label{fig:wrapJ}} {
    	\vspace{0.5em}
    	\includegraphics[height=1.0in,trim={0.3in 0.2in 0.9in 0.8in},clip]{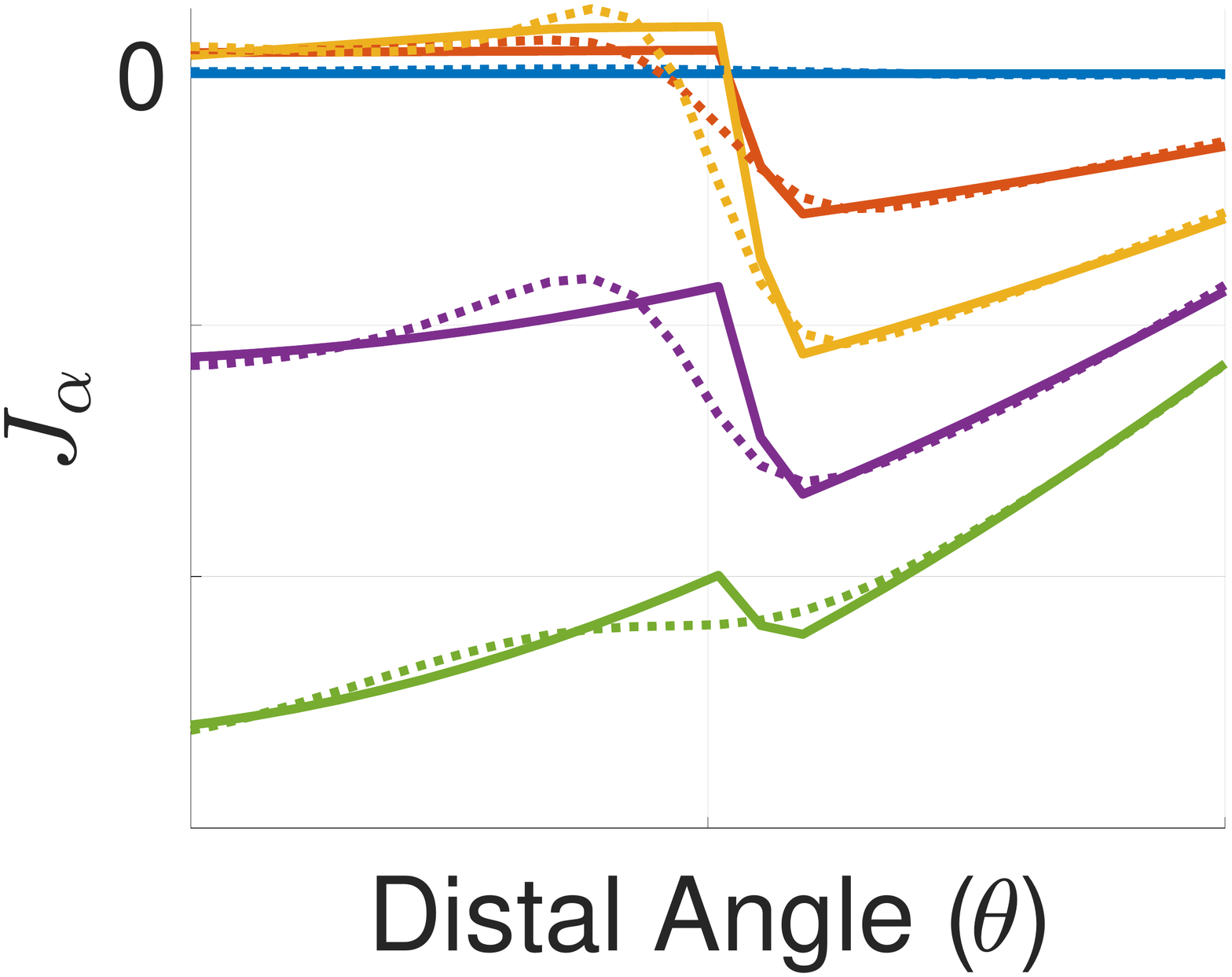}
    }
    \vspace{-0.1in}
    \caption{
      (a) A double pendulum with a musculotendon, hitting a wrapping surface.
      (b) Energy plot (kinetic in {\color{MATBLUE}blue}, potential in {\color{MATRED}red}, total in {\color{MATYELLOW}yellow}) of the simulation using an existing wrapping surface library.
      (c) Energy plot using our approach.
      (d) Plot of the x-component of \edit{five selected} muscle mass points as a function of the distal joint angle, zoomed around a discontinuity.
      The solid lines are generated using an existing wrapping surface library.
      The dotted lines are generated using our approach.
      (e) The corresponding plots of the Jacobian.
      Unlike previous work (solid), our approach (dotted) generates smooth Jacobians.
    }
    \label{fig:wrapCollision}
  \end{figure*}
}

\newcommand{\figEOLbeta}{
  \begin{figure}[b!]
    \centering
    \subcaptionbox{\label{fig:EOL}} {
	    \includegraphics[width=0.26\columnwidth,trim={0in 0in 0in 0in},clip]{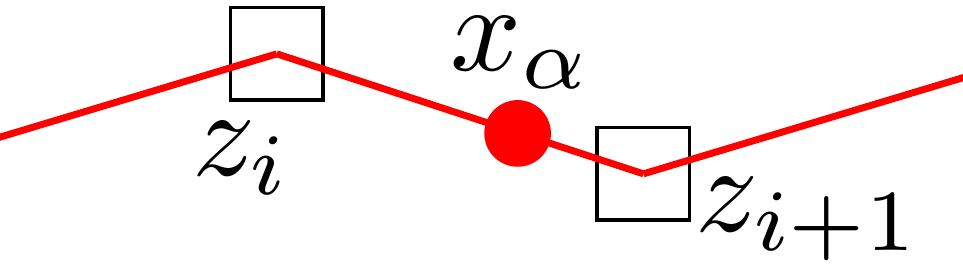}
	}
	\hspace{0.5in}
	\subcaptionbox{\label{fig:beta}} {
	    \includegraphics[width=0.26\columnwidth,trim={0in 0in 0in 0in},clip]{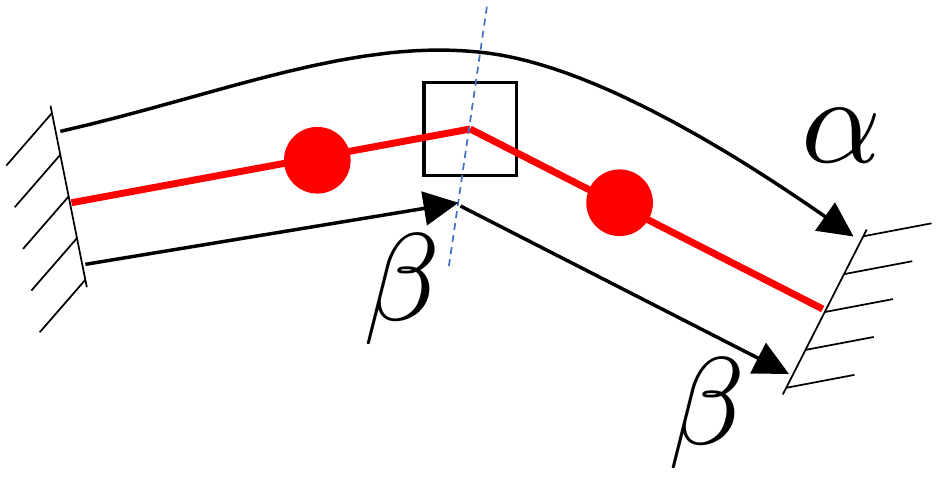}
	    \vspace{-0.1in}
	}
  \vspace{-0.1in}
    \caption{
      (a) An EOL segment: the motion of the mass point $\xx_\alpha$ depends on the motion of both Eulerian and Lagrangian motions of the path points $\zz_i$ and $\zz_{i+1}$.
      (b) A musculotendon with one path point between origin and insertion: $\alpha$ represents the percentage length along the whole musculotendon, whereas $\beta$ represents the percentage length along each line segment.
    }
    \label{fig:EOLbeta}
  \end{figure}
}

\newcommand{\figWrapSpaces}{
  \begin{figure}[t]
    \centering
    \includegraphics[width=0.5\columnwidth,trim={0in 0in 0in 0in},clip]{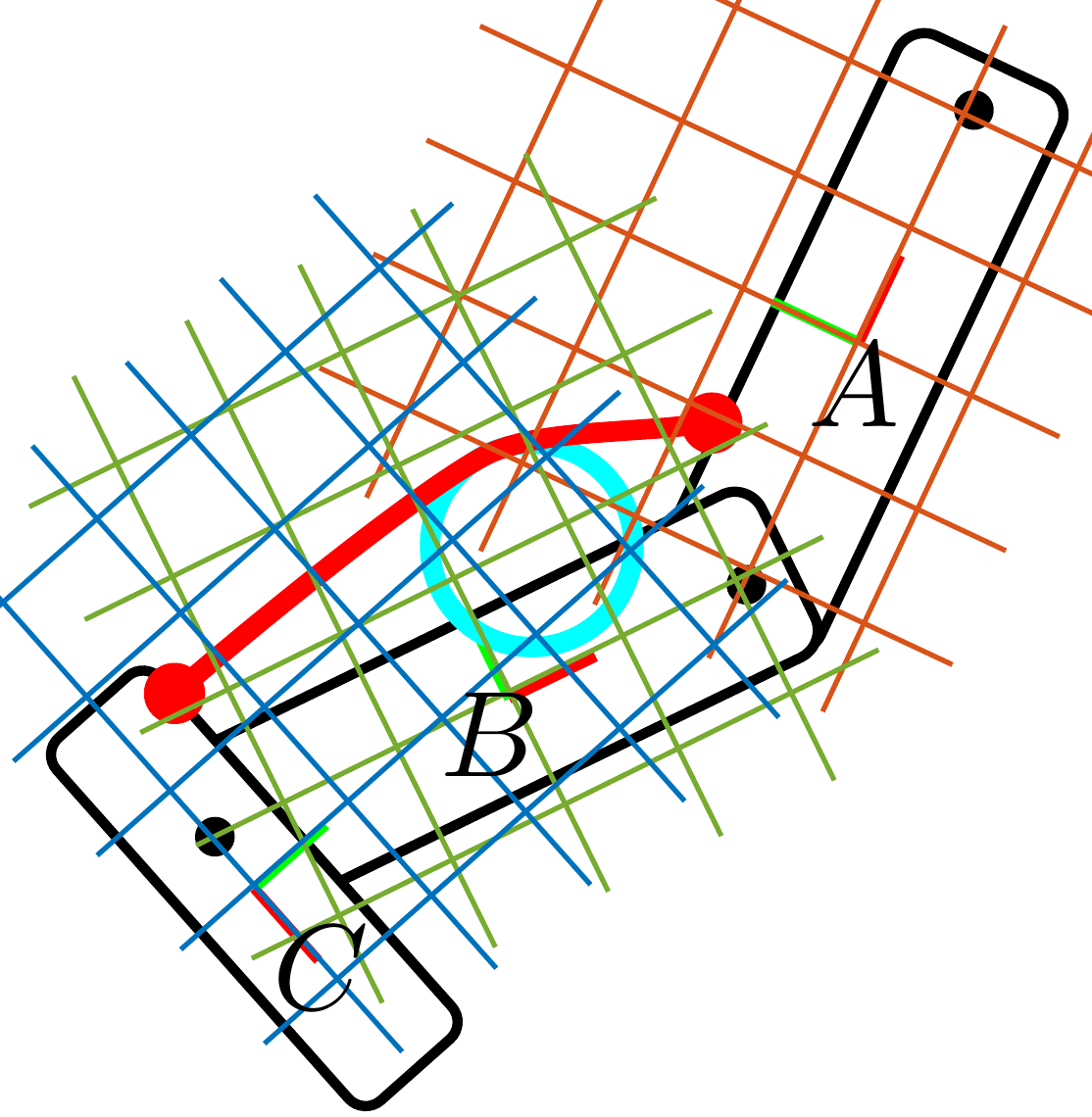}
    \caption{
      Coordinate spaces for a wrapping surface muscle. $A$ contains the origin, $C$ contains the insertion, and $B$ contains the wrapping surface $S$.
      The $S$ coordinate space (not drawn in this figure) moves rigidly with $B$.
    }
    \label{fig:wrapSpaces}
  \end{figure}
}

\newcommand{\figDinesh}{
  \begin{figure}[b]
  	\captionsetup[subfigure]{}
    \centering
    \subcaptionbox{\label{fig:dineshSim}} {
    	\includegraphics[height=1.1in,trim={0in 0in 0in 0in},clip]{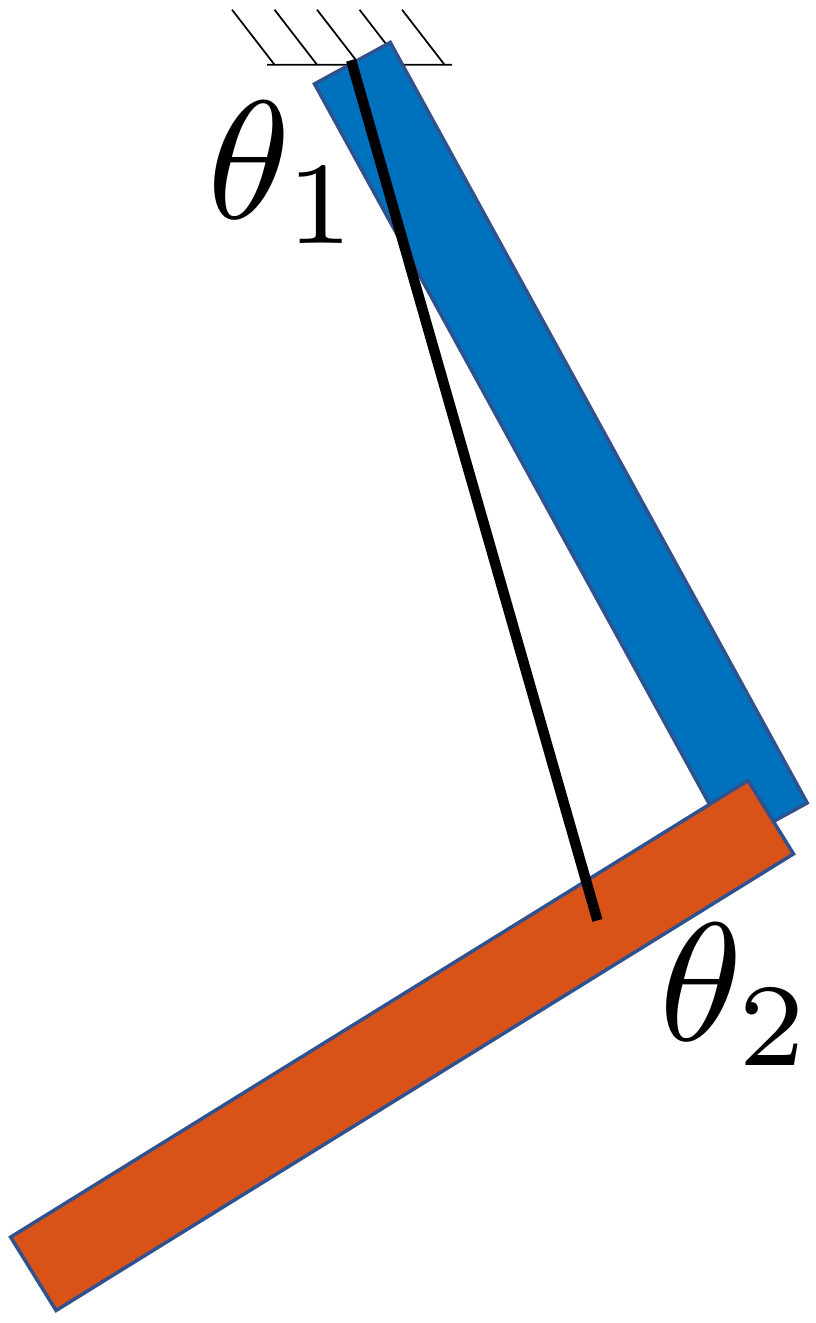}
    }
    \subcaptionbox{\label{fig:dineshPlot}} {
    	\includegraphics[height=1.1in,trim={0in 0in 0in 0in},clip]{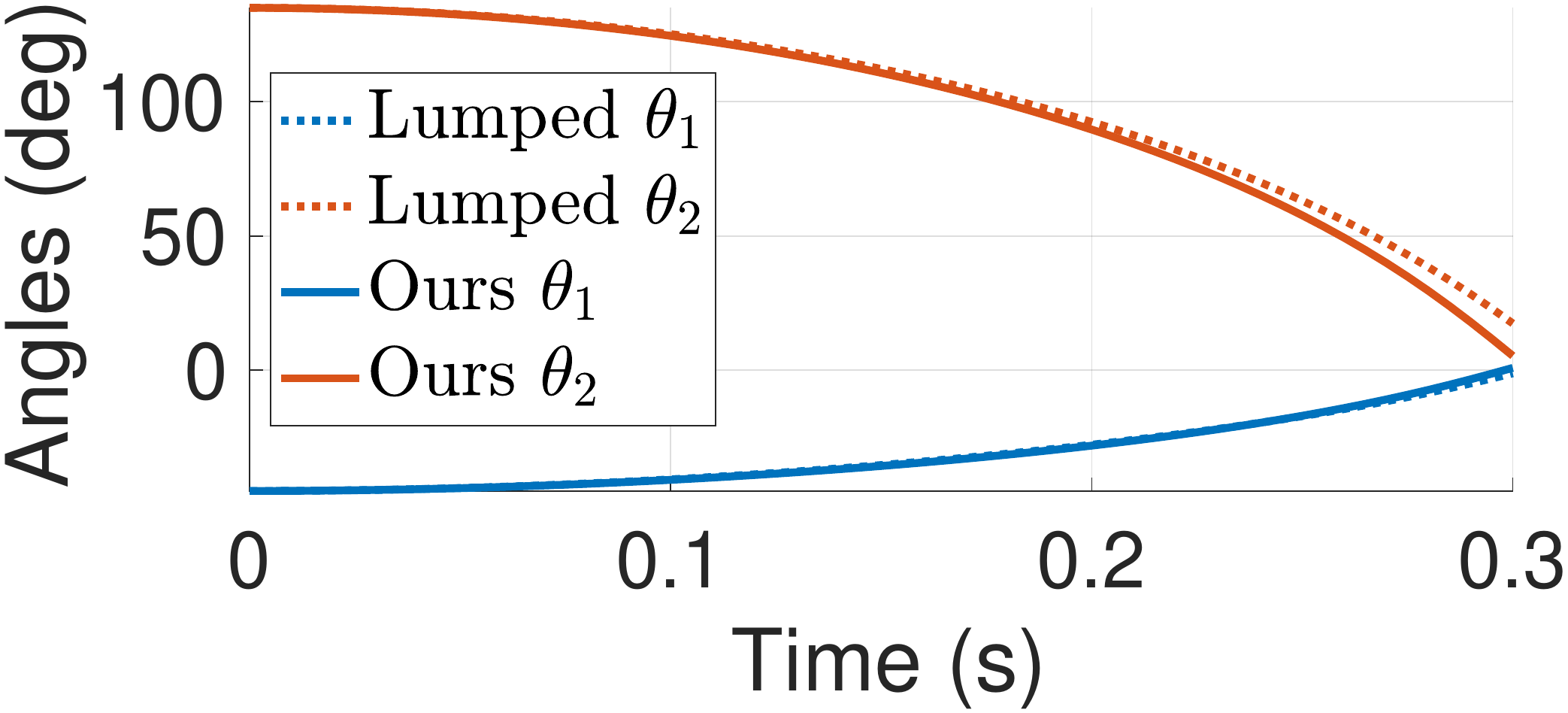}
    }
    \caption{
      Comparison to published results \cite{Pai2010}.
      (a) Two bones and one muscle, all with the same mass.
      (b) The solid lines show that after simulating the system with the muscle for 0.3 seconds, the two angles straighten out as in the previous work.
      The dotted lines show the same simulation but with the mass of the muscle lumped onto the bones.
    }
    \label{fig:dinesh}
  \end{figure}
}

\newcommand{\figCylinders}{
  \begin{figure}[t]
    \centering
    \includegraphics[width=0.99\columnwidth,trim={0in 0in 0in 0in},clip]{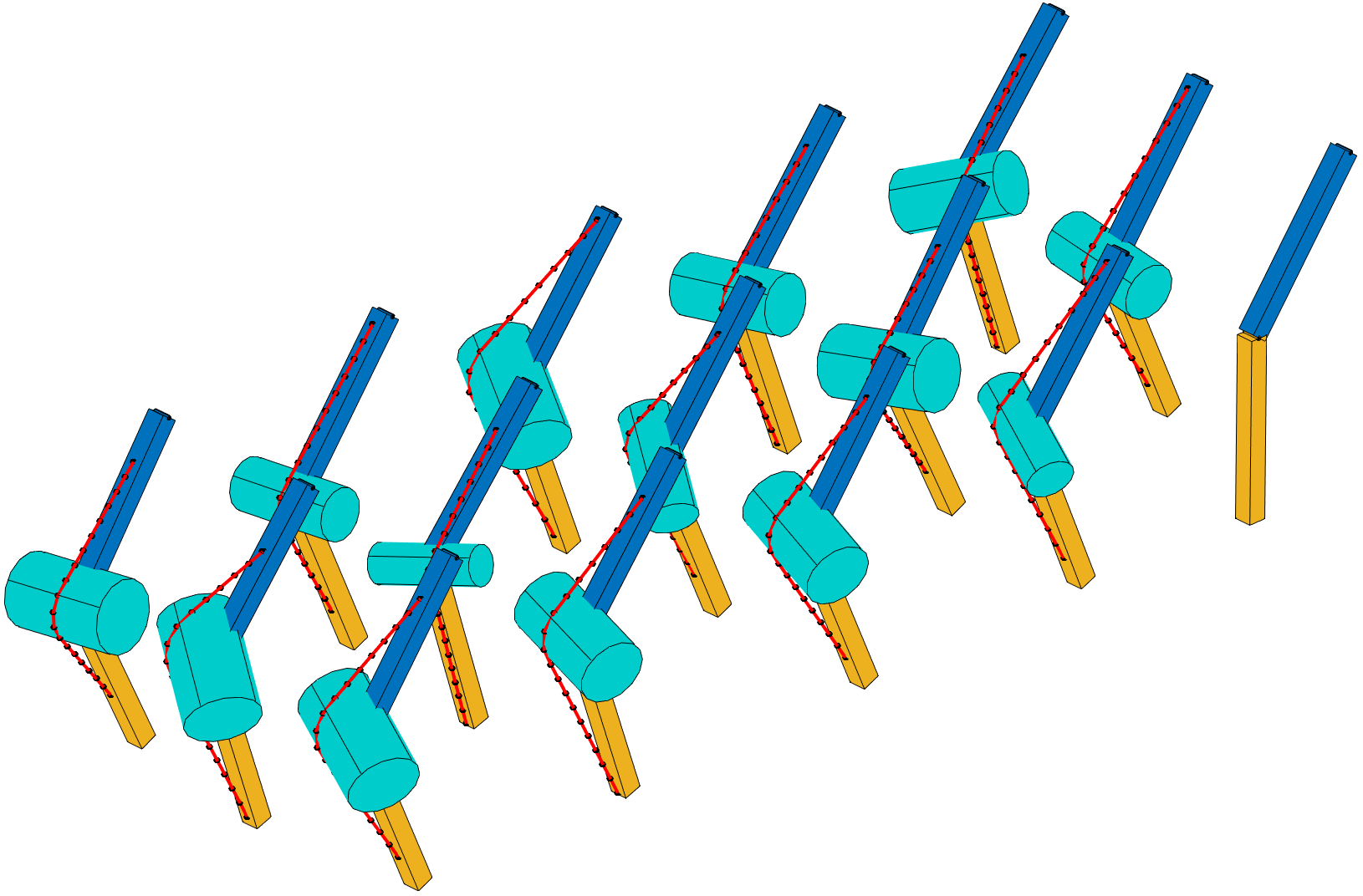}
    \caption{
      Double pendulums with cylinder wrapping.
      The same trained network is used for a range of input parameters.
      For comparison, the right-most double pendulum is simulated without a muscle.
    }
    \vspace{-0.1in}
    \label{fig:cylinders}
  \end{figure}
}

\newcommand{\figQvv}{
  \begin{figure}[t]
    \centering
    \subcaptionbox{\label{fig:QvvOffTypeII}\editTwo{Type II without QVV}} {
    	\includegraphics[width=0.47\columnwidth,trim={0in 0in 0in 0.0in},clip]{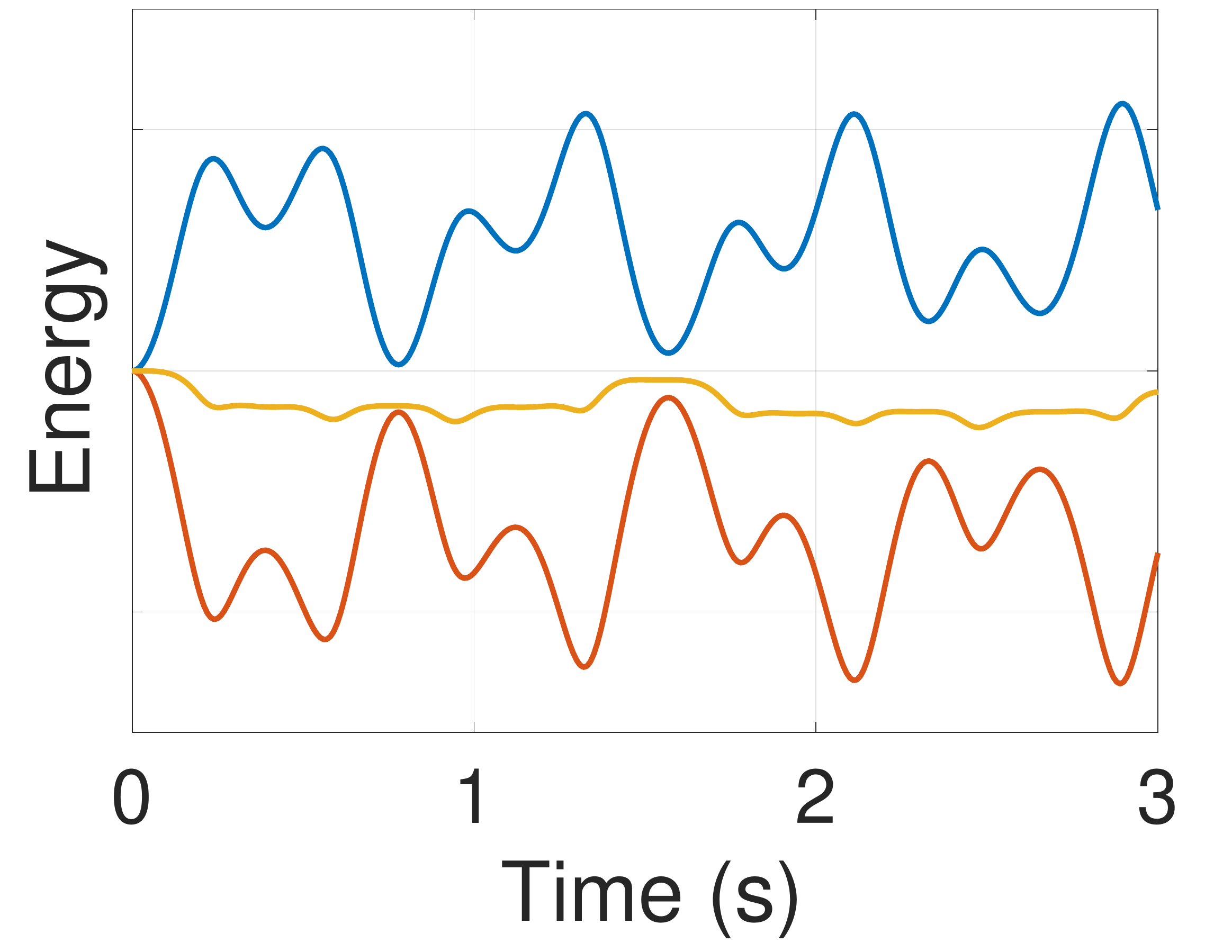}
    }
    \subcaptionbox{\label{fig:QvvOnTypeII}\editTwo{Type II with QVV}} {
    	\includegraphics[width=0.47\columnwidth,trim={0in 0in 0in 0.0in},clip]{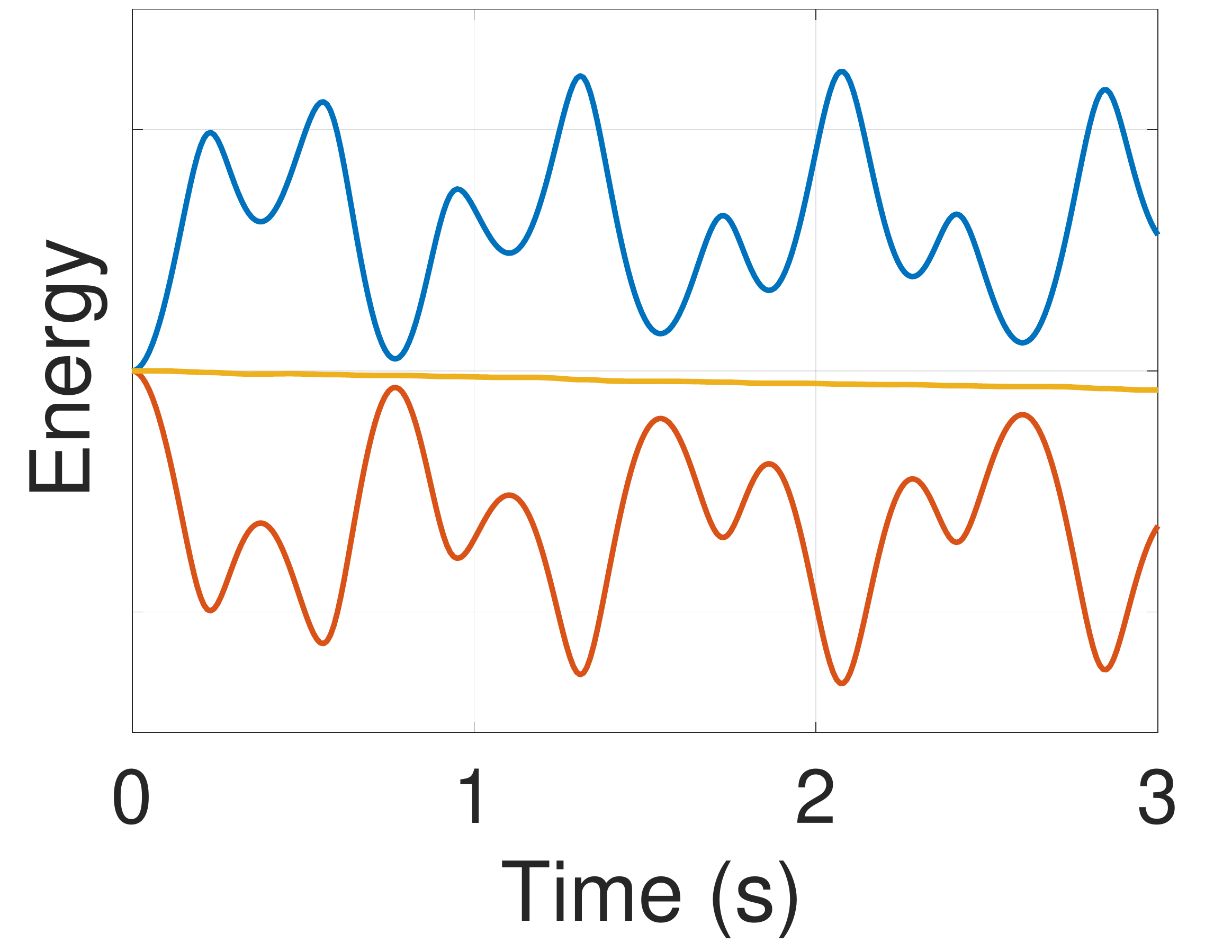}
    }
	\\\vspace{0.2in}
    \subcaptionbox{\label{fig:QvvOffTypeIII}\editTwo{Type III without QVV}} {
    	\includegraphics[width=0.47\columnwidth,trim={0in 0in 0in 0.0in},clip]{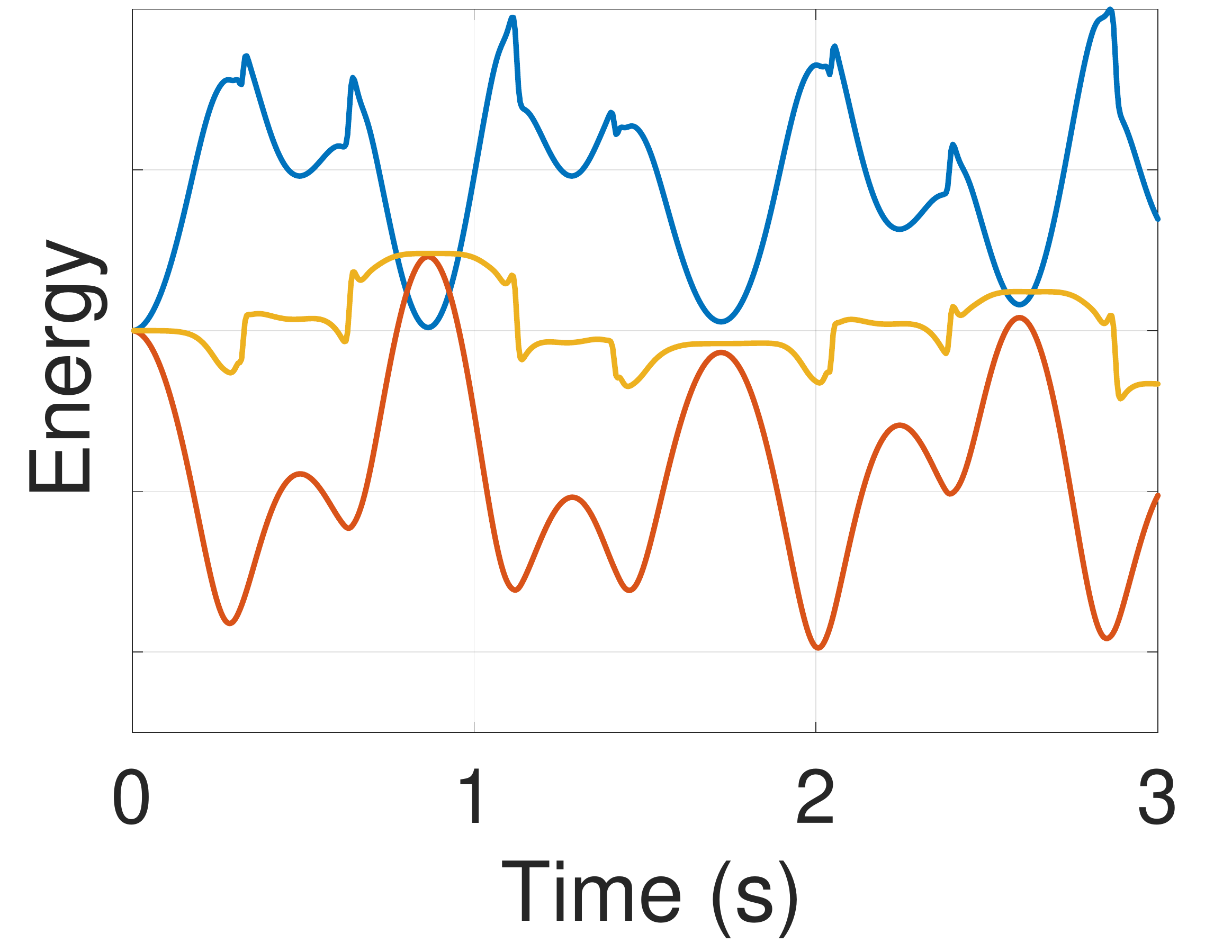}
    }
    \subcaptionbox{\label{fig:QvvOnTypeIII}\editTwo{Type III with QVV}} {
    	\includegraphics[width=0.47\columnwidth,trim={0in 0in 0in 0.0in},clip]{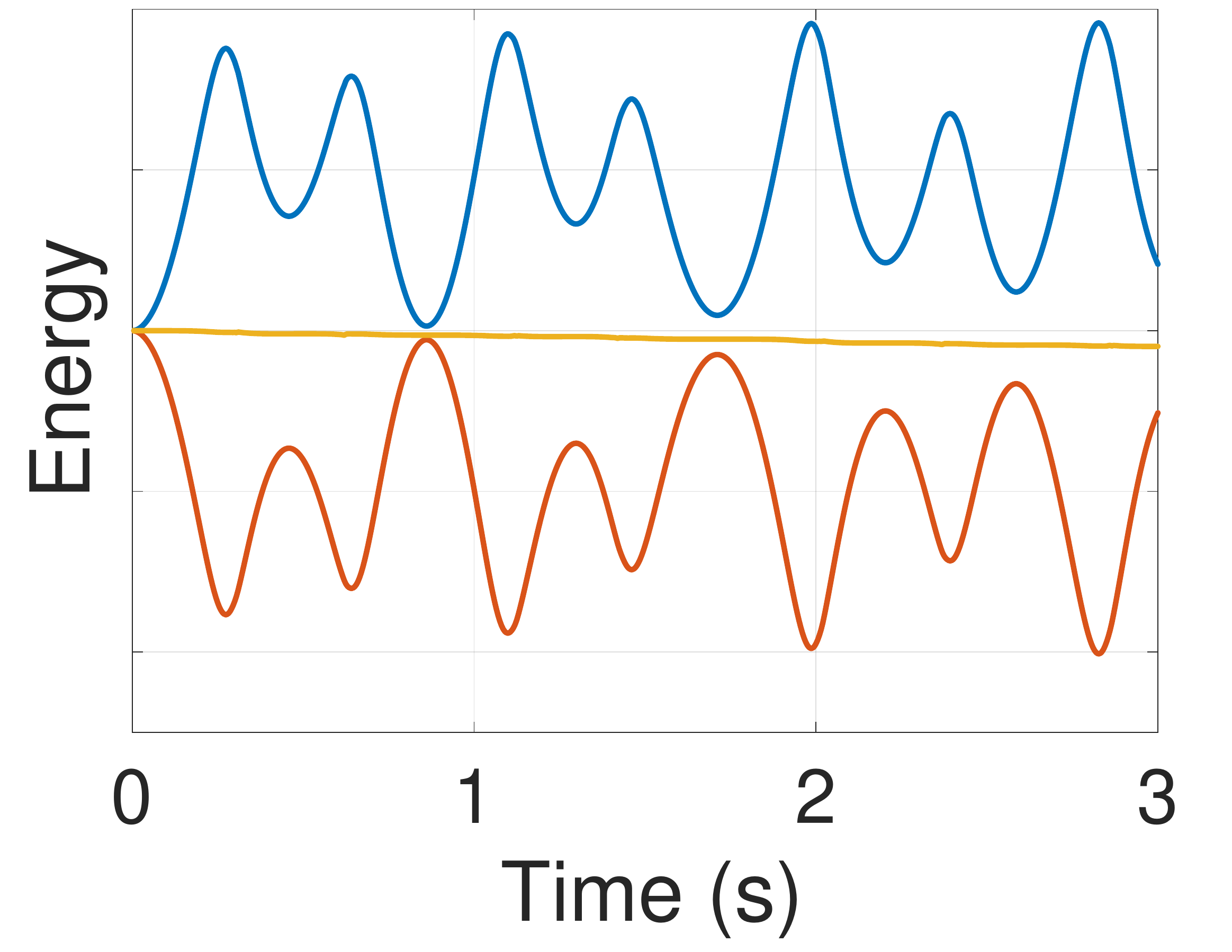}
    }
    \caption{
      (a-b) Energy plots from a Type II muscle with and without QVV.
      (c-d) Energy plots from a Type III muscle with and without QVV.
      Kinetic energy is shown in {\color{MATBLUE}blue}, potential energy in {\color{MATRED}red}, and total energy in {\color{MATYELLOW}yellow}.
    }
    \label{fig:Qvv}
  \end{figure}
}

\newcommand{\figRun}{
  \begin{figure*}[t]
  	\captionsetup[subfigure]{labelformat=empty}
    \centering
    \subcaptionbox{$t=0.0$} {
    	\includegraphics[width=0.154\textwidth,trim={6.1in 7.3in 7in 5.3in},clip]{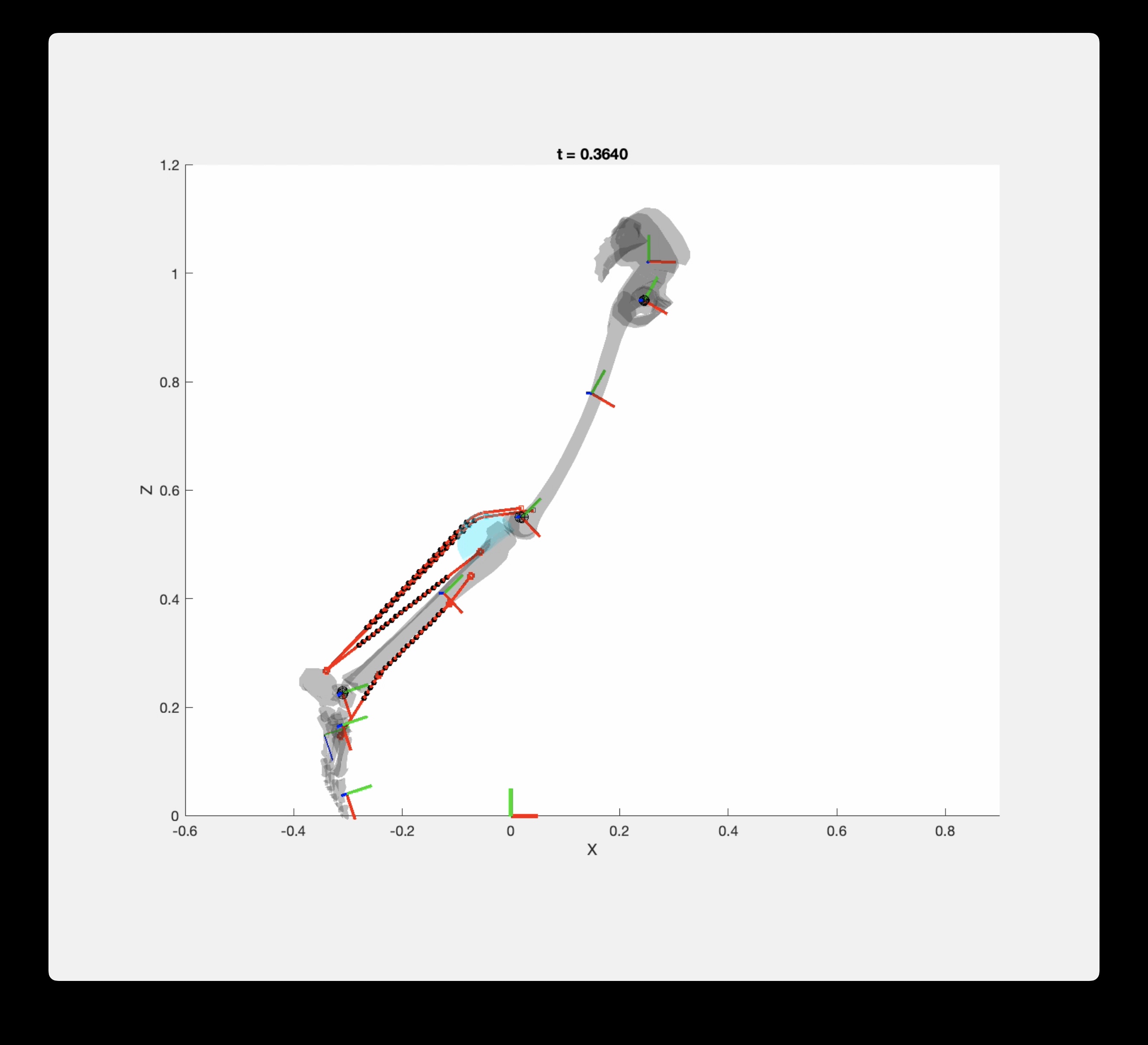}
    }
    \subcaptionbox{$t=0.1$} {
    	\includegraphics[width=0.154\textwidth,trim={6.1in 7.3in 7in 5.3in},clip]{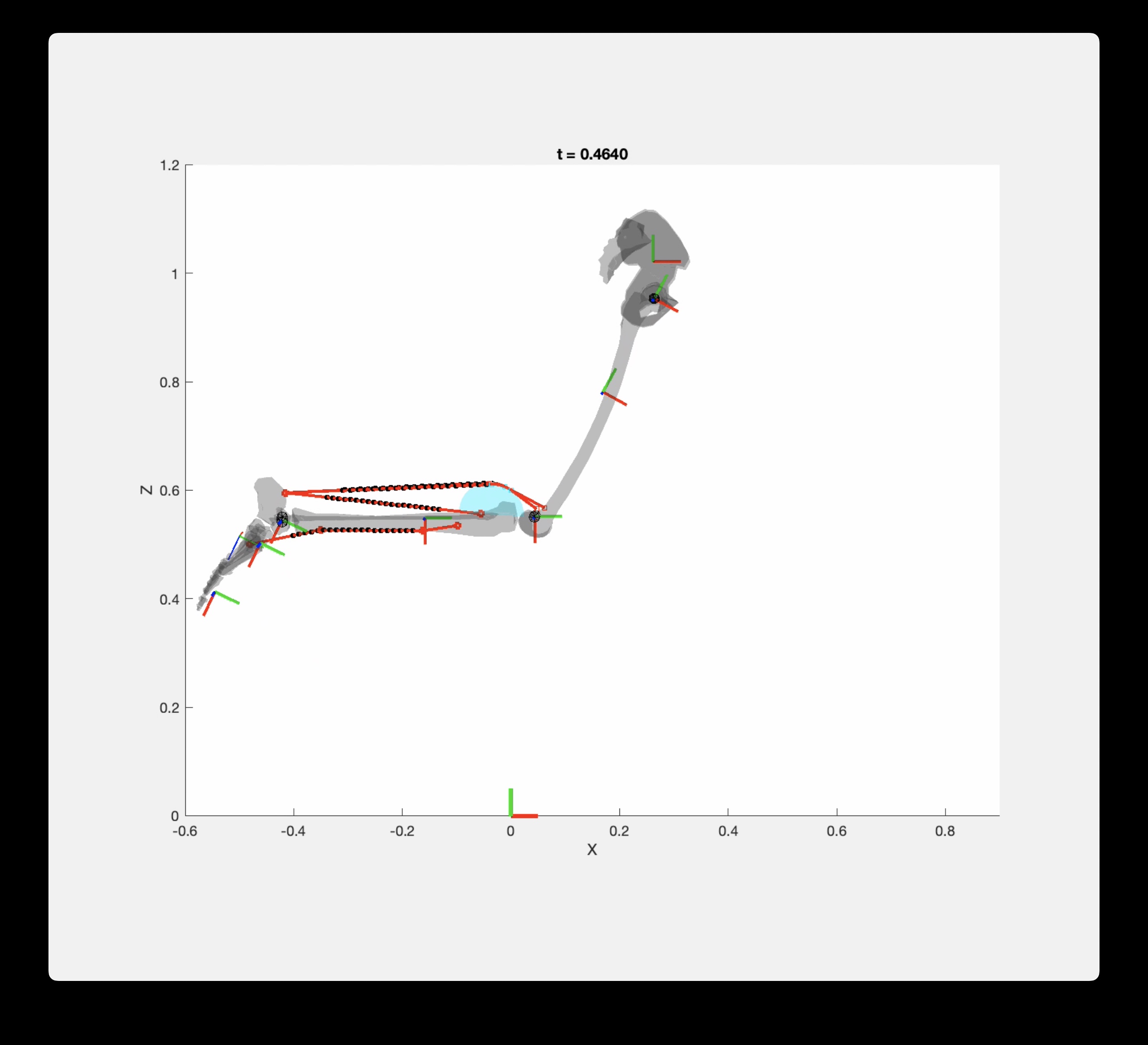}
    }
    \subcaptionbox{$t=0.2$} {
    	\includegraphics[width=0.154\textwidth,trim={6.1in 7.3in 7in 5.3in},clip]{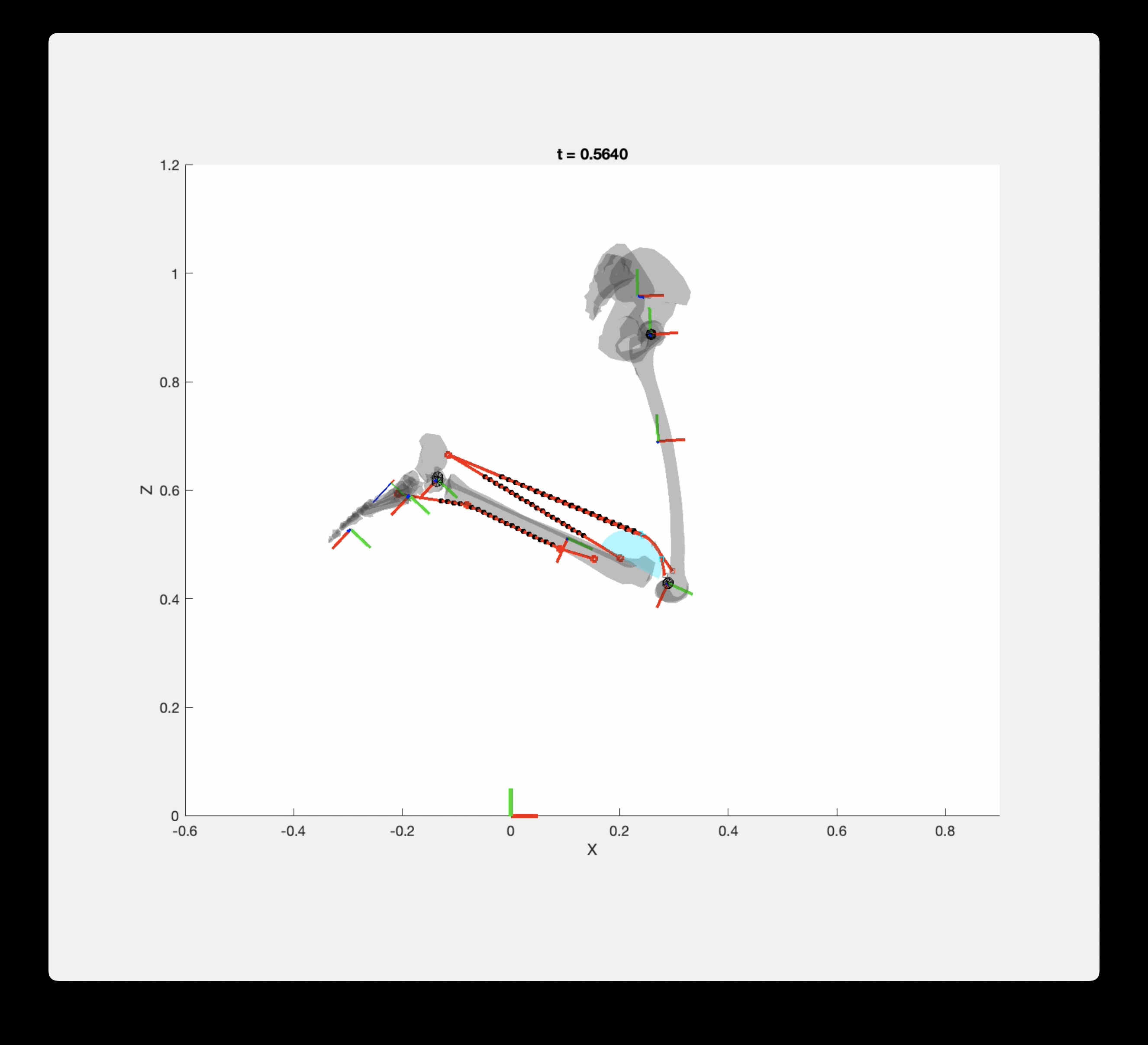}
    }
    \subcaptionbox{$t=0.3$} {
    	\includegraphics[width=0.154\textwidth,trim={6.1in 7.3in 7in 5.3in},clip]{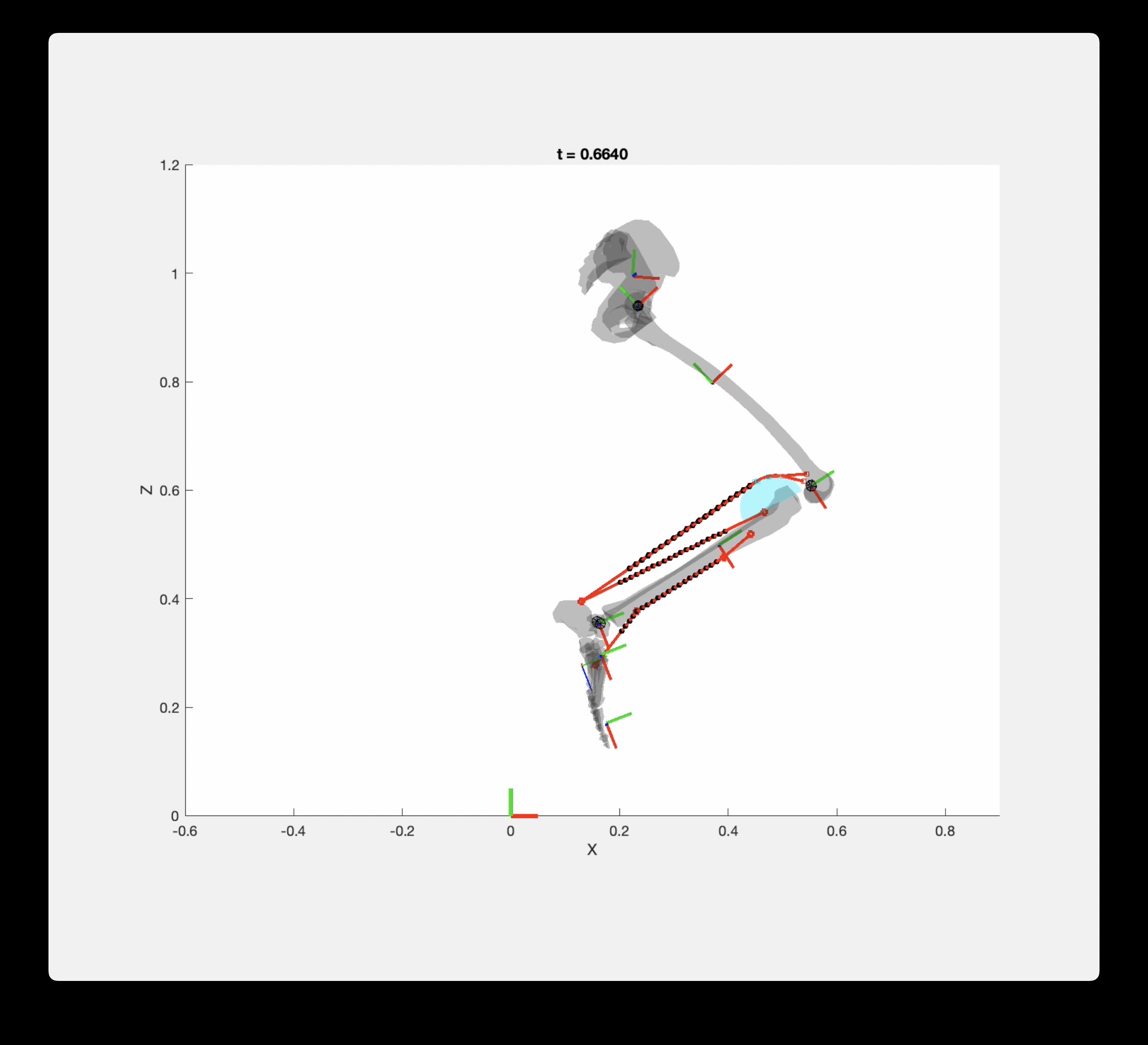}
    }
    \subcaptionbox{$t=0.4$} {
    	\includegraphics[width=0.154\textwidth,trim={6.1in 7.3in 7in 5.3in},clip]{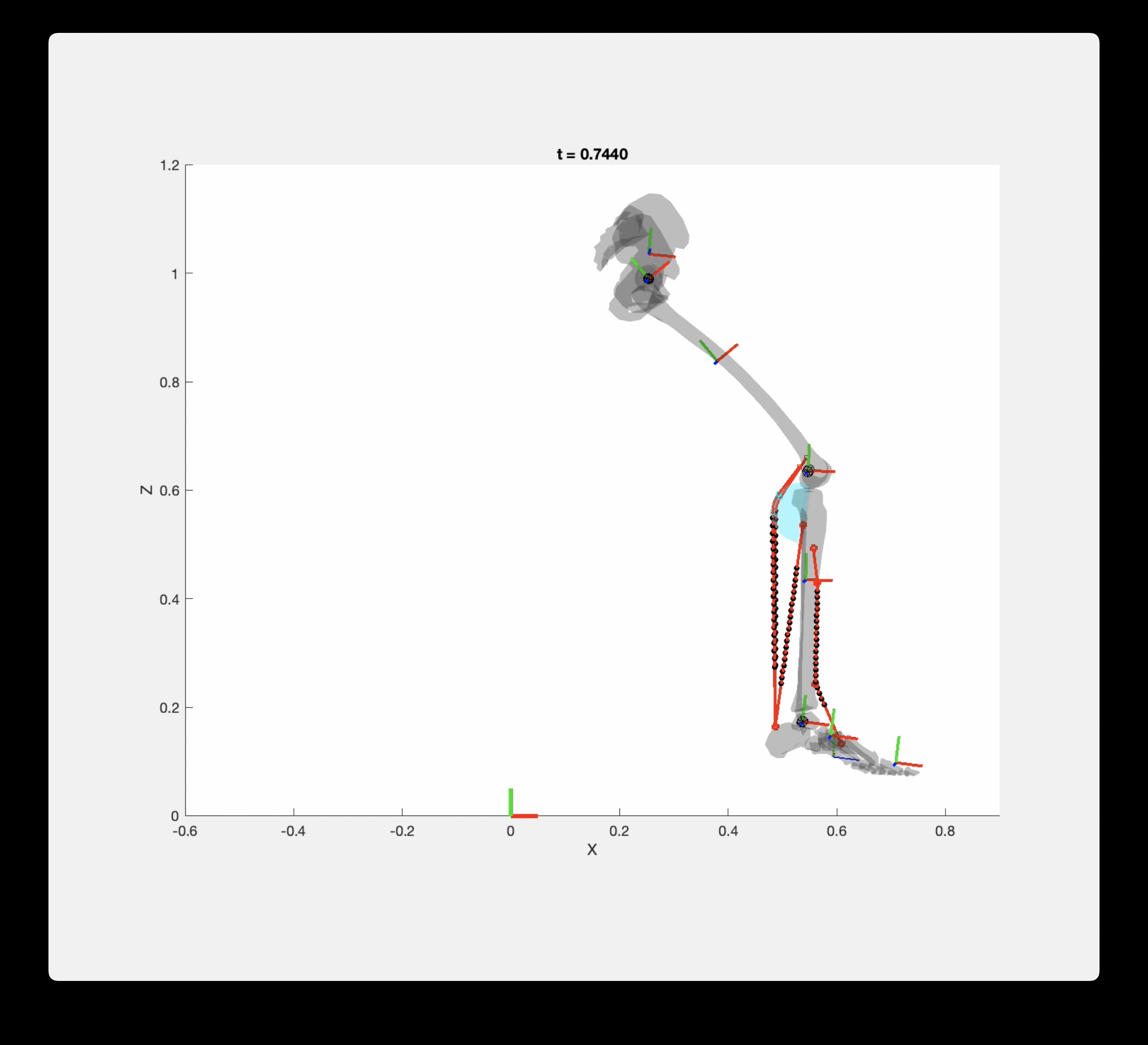}
    }
    \subcaptionbox{$t=0.5$} {
    	\includegraphics[width=0.154\textwidth,trim={6.1in 7.3in 7in 5.3in},clip]{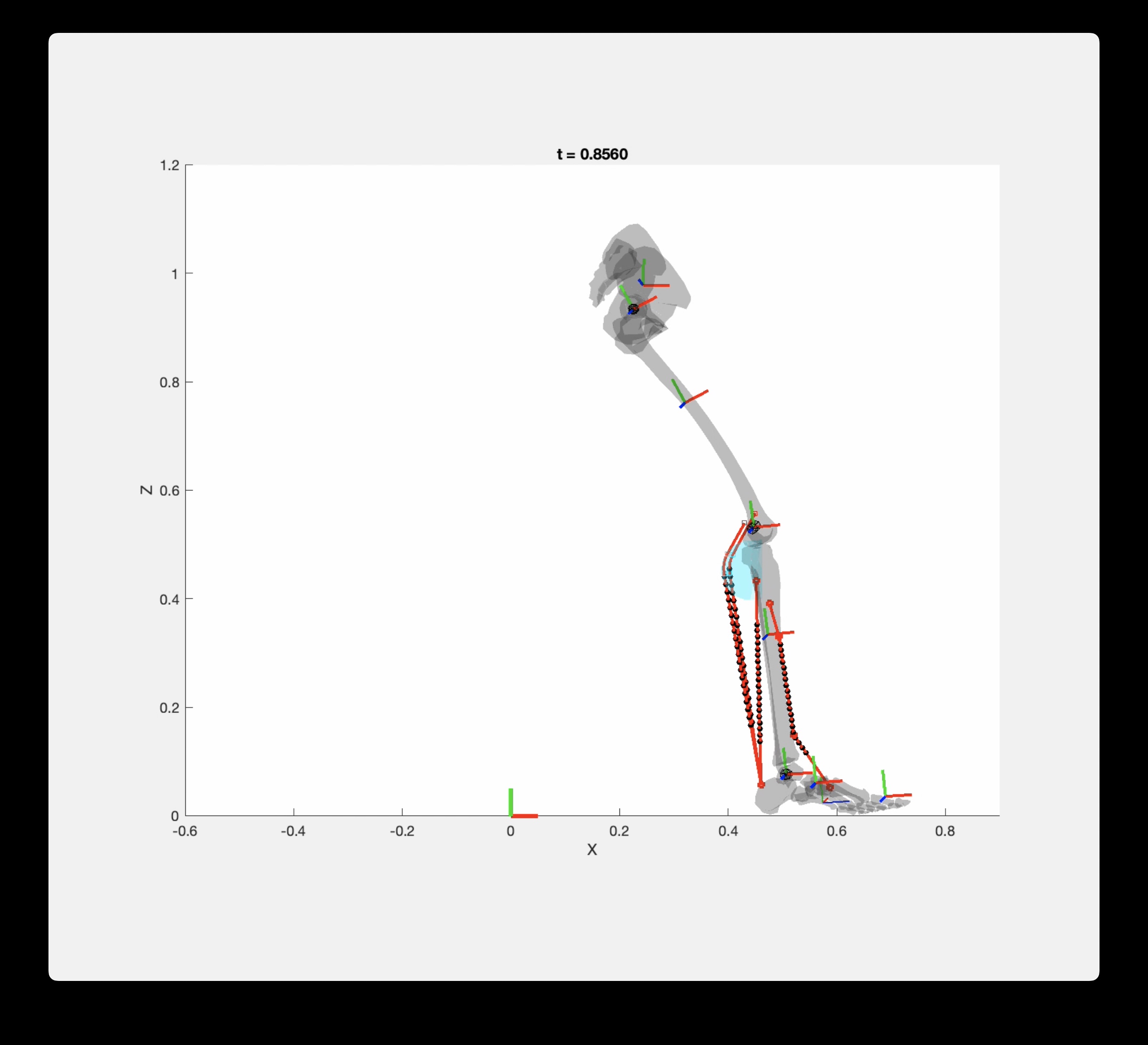}
    }
    \vspace{-0.1in}
    \caption{
      The swing phase of a \editTwo{19.1 km/h} treadmill run, showing only the right leg.
      The four muscles (and their types) are: gastrocnemius lateral (Type III), gastrocnemius medial (Type III), soleus (Type I), and tibialis anterior (Type II).
    }
    \label{fig:run}
  \end{figure*}
}

\newcommand{\figTorques}{
  \begin{figure}[t]
  	\captionsetup[subfigure]{labelformat=empty}
    \centering
    \subcaptionbox{\label{fig:torques19.1}} {
    	\includegraphics[height=1.3in,trim={0.15in 0.0in 1.6in 0.45in},clip]{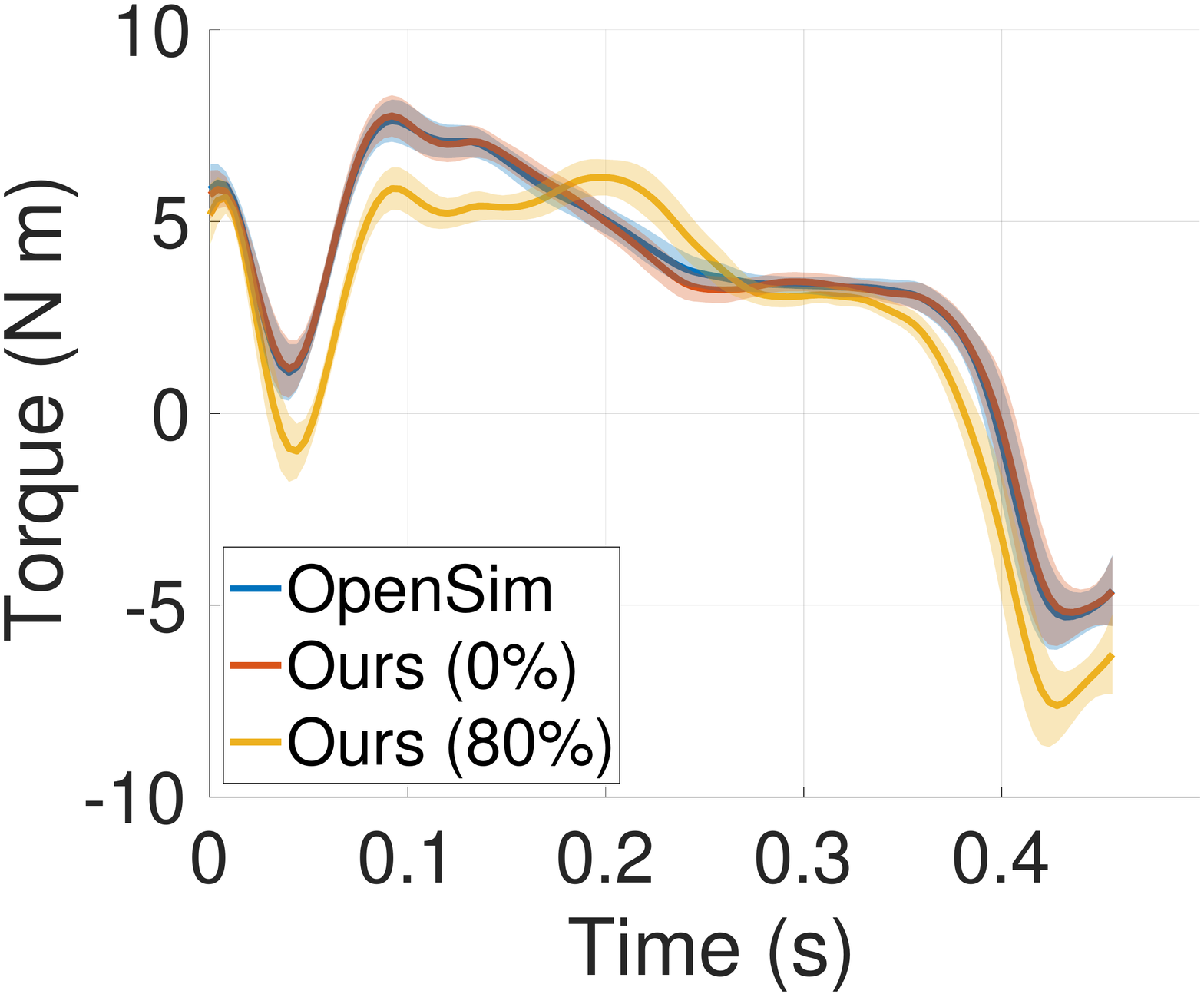}
    }
    \subcaptionbox{\label{fig:torques19.1zoom}} {
    	\includegraphics[height=1.3in,trim={0.8in 0.0in 0.5in 0.5in},clip]{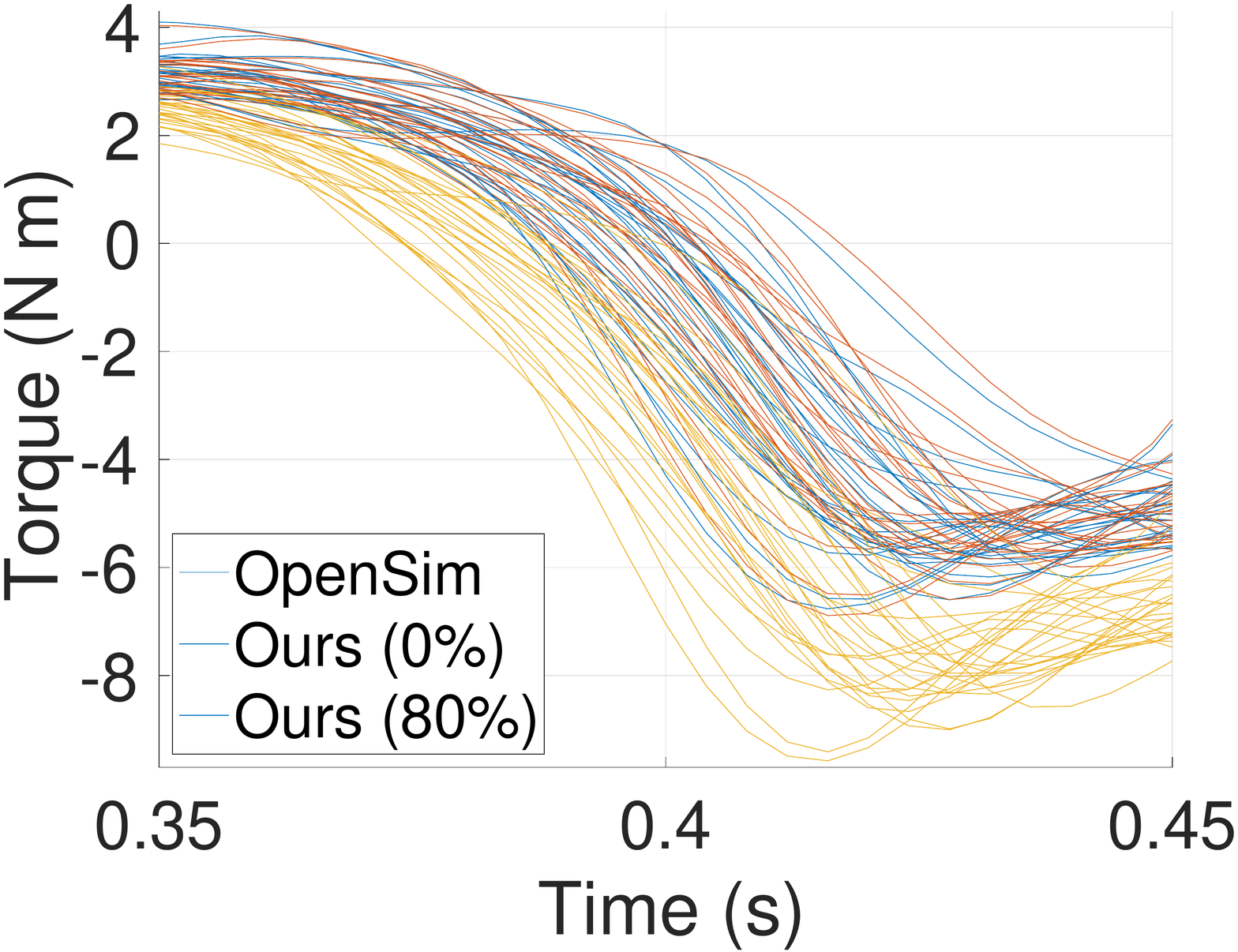}
    }
    \vspace{-0.3in}
    \caption{
      \editTwo{(Left) Ankle torque computed by inverse dynamics, \edit{showing the mean and the standard deviation.}}
      {\color{MATBLUE}Blue} plot is generated by OpenSim, which does not support inertial muscles.
      {\color{MATRED}Red} plot is generated by our simulator with the muscles accounting for 0\% of the total mass.
      {\color{MATYELLOW}Yellow} plot is generated by our simulator with 80\% of the tibia segment mass transferred to the muscles.
      (Right) The closeup of the final dip, \edit{showing the individual trajectories}.
      Our simulator generates results that gracefully degrade to OpenSim's results as the inertia of the muscles is decreased to zero.
    }
    \label{fig:torques}
  \end{figure}
}

\newcommand{\figKnee}{
  \begin{figure}[t]
    \centering
    \includegraphics[width=0.99\columnwidth,trim={0in 0.0in 0in 0in},clip]{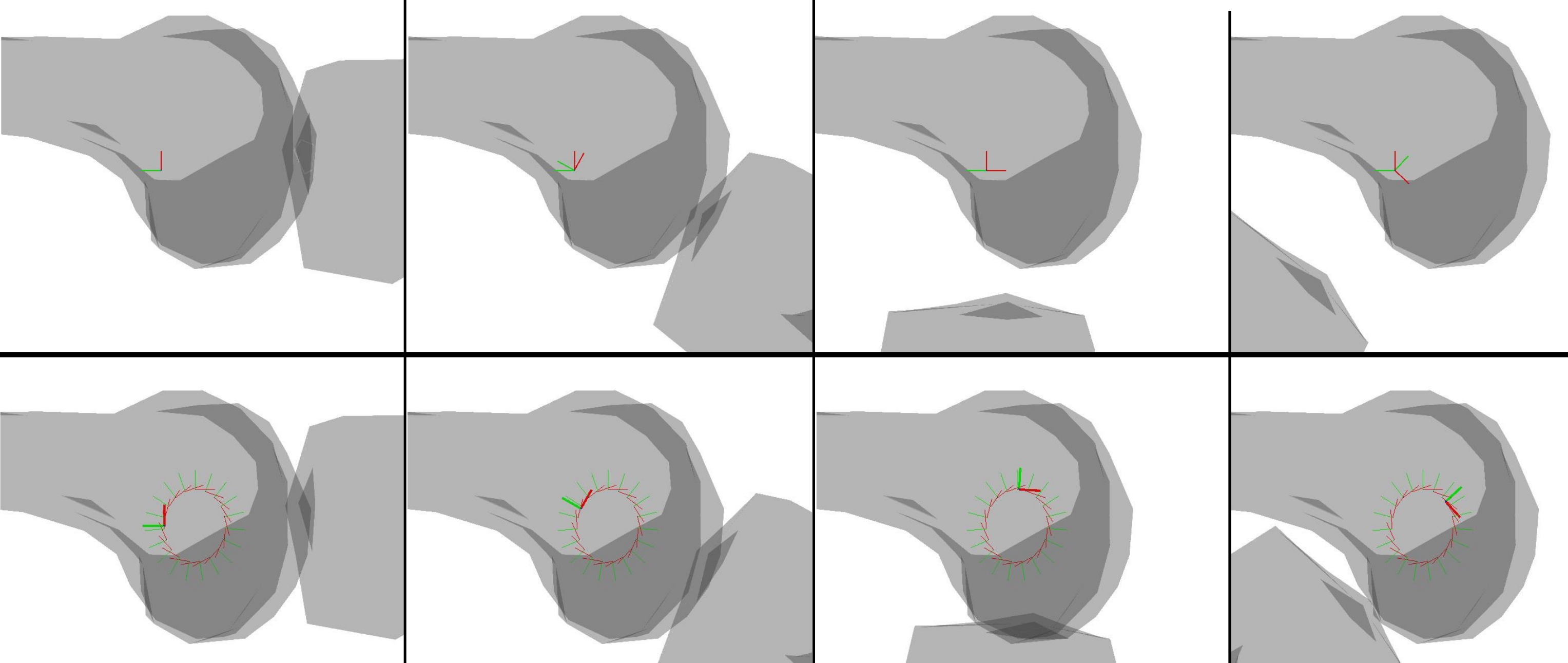}
    \caption{
      Our approach supports complex joint types.
      Top row: Knee with a revolute joint---the tibia separates from the femur.
      Bottom row: Knee with a spline joint---the tibia stays close to the femur.
    }
    \label{fig:knee}
  \end{figure}
}

\newcommand{\figKneeAnim}{
  \begin{figure*}[tb]
  	\captionsetup[subfigure]{labelformat=empty}
    \centering
    \subcaptionbox{} {
    	\includegraphics[height=1in,trim={2.6in 6.6in 4.8in 1.9in},clip]{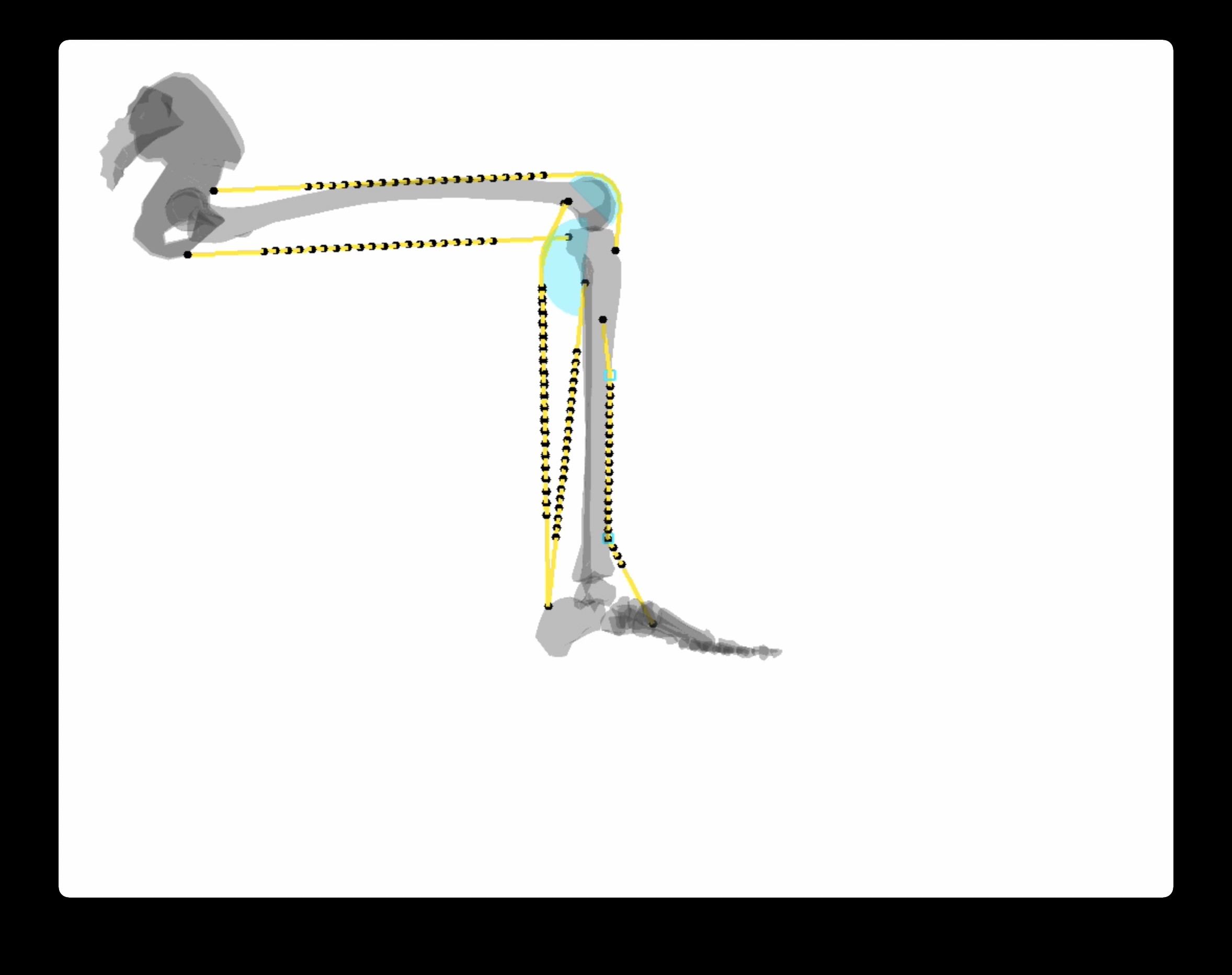}
    }
    \subcaptionbox{} {
    	\includegraphics[height=1in,trim={2.6in 6.6in 4.8in 1.9in},clip]{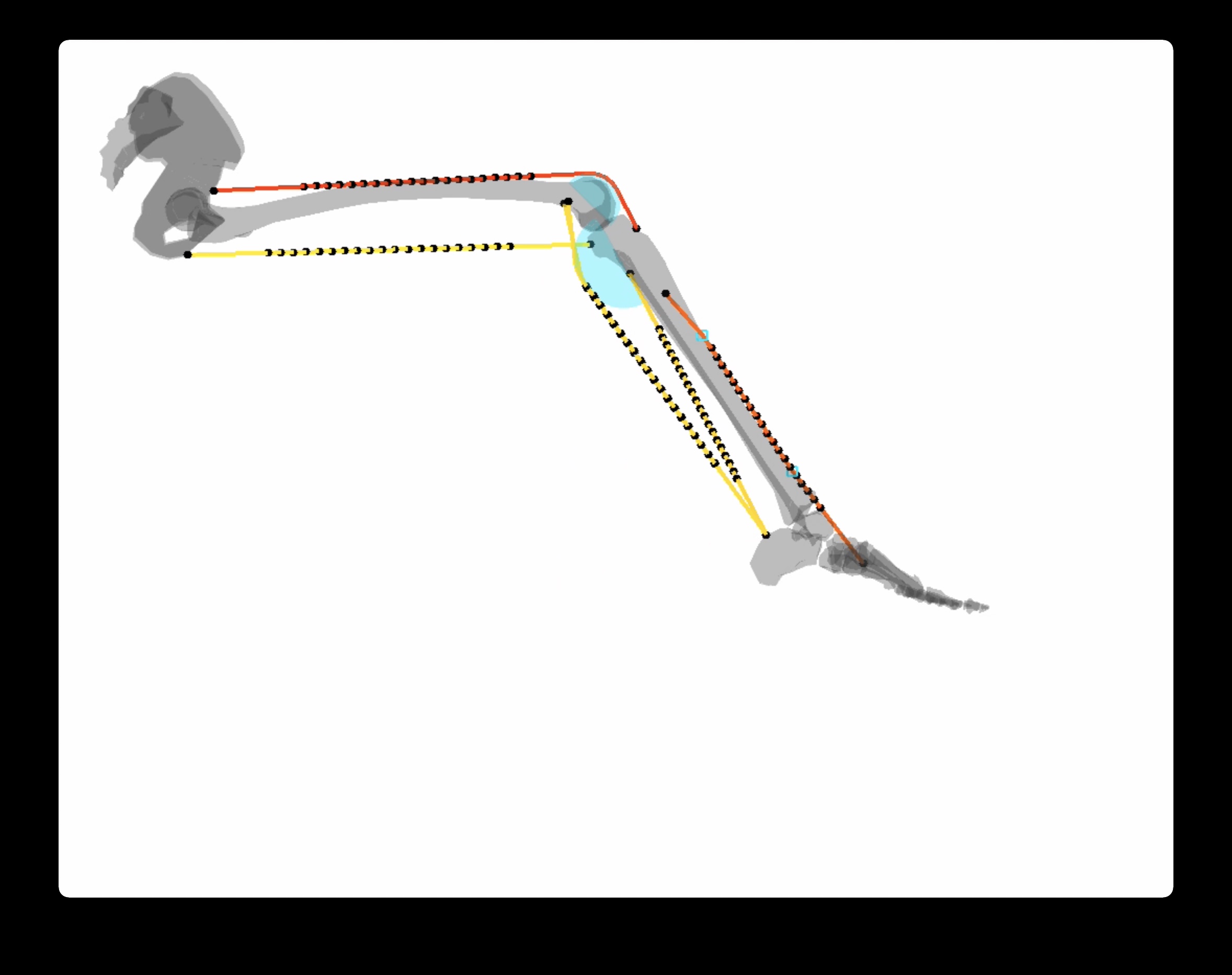}
    }
    \subcaptionbox{} {
    	\includegraphics[height=1in,trim={2.6in 6.6in 4.8in 1.9in},clip]{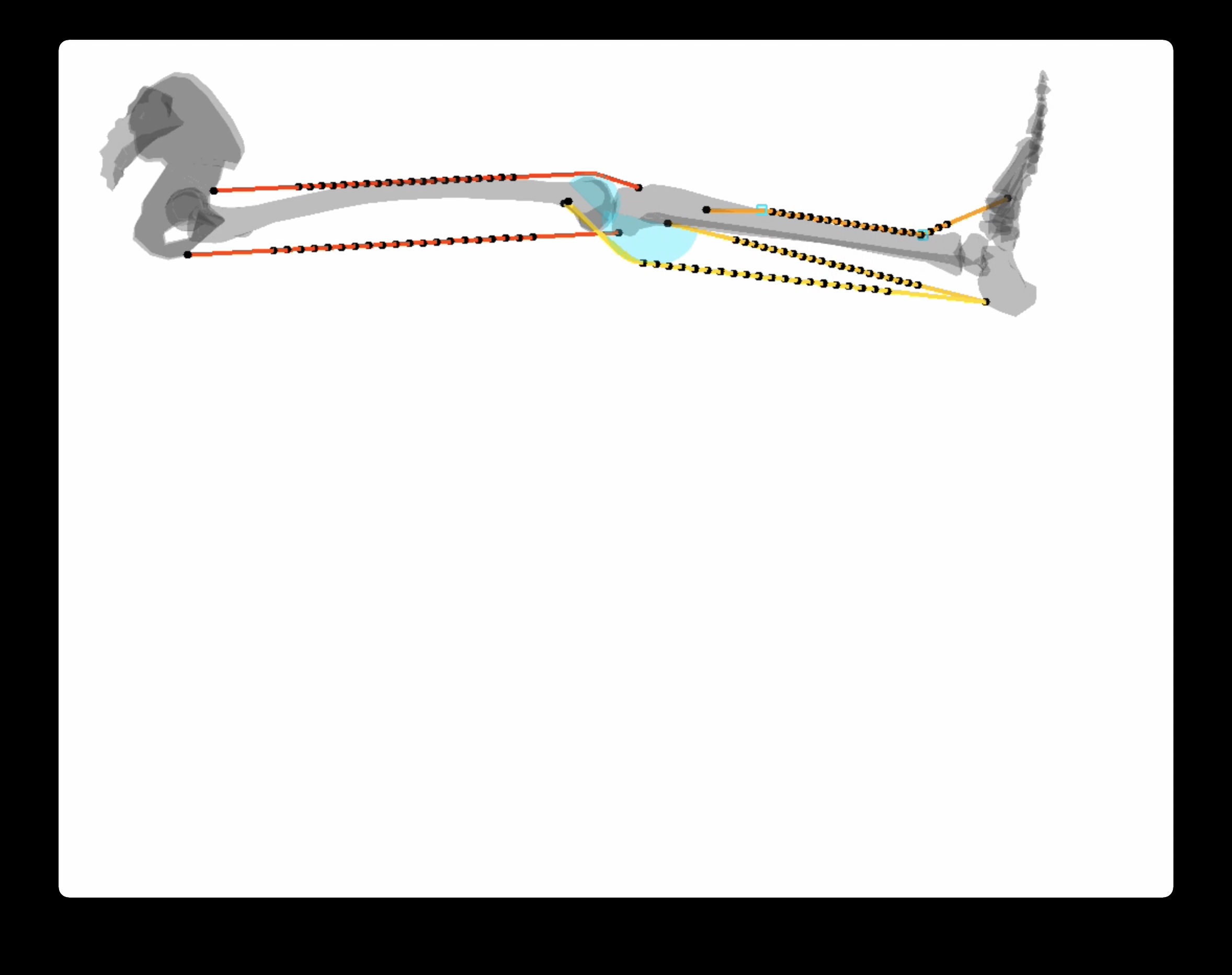}
    }
    \subcaptionbox{} {
    	\includegraphics[height=1in,trim={2.6in 6.6in 4.8in 1.9in},clip]{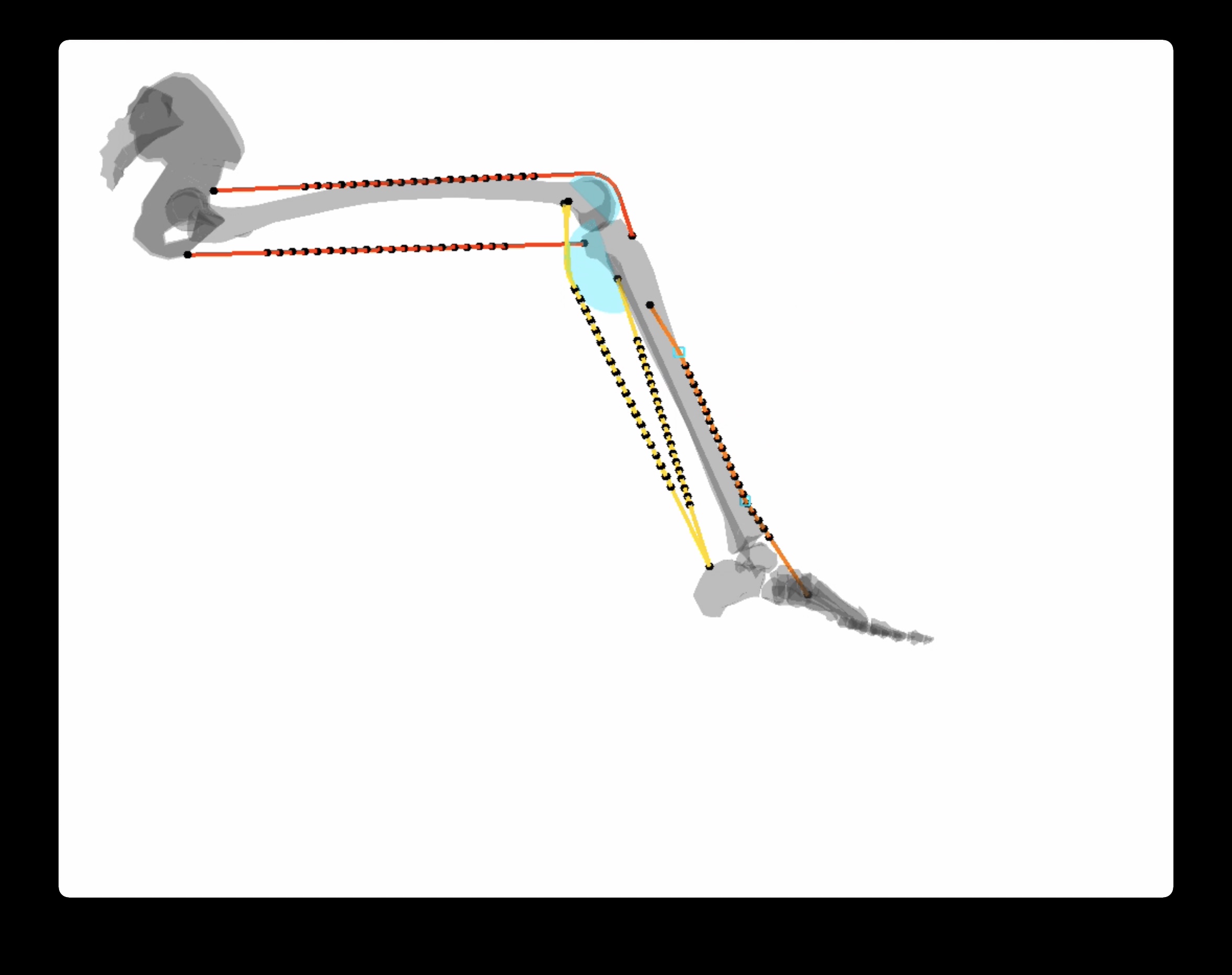}
    }
    \subcaptionbox{} {
    	\includegraphics[height=1in,trim={2.6in 6.6in 16.2in 1.9in},clip]{figs/kneeAnim4.jpg}
    }
    \vspace{-0.3in}
    \caption{
      For \S\ref{sec:splineHill}, we add a spline joint knee and Hill-type muscles to the model used in \S\ref{sec:comparison}.
      We manually excite the rectus femoris and the semimembranosus muscles.
      The excitation levels of the soleus and the tibialis anterior muscles are computed automatically with a proportional controller.
    }
    \label{fig:kneeAnim}
  \end{figure*}
}

\setcitestyle{nosort,square} 

\setcopyright{acmcopyright}
\acmJournal{TOG}
\acmYear{2022}
\acmVolume{41}
\acmNumber{6}
\acmArticle{272}
\acmMonth{12}
\acmDOI{10.1145/3550454.3555490}

\begin{document}
\title{Differentiable Simulation of Inertial Musculotendons}


\author{Ying Wang}
\orcid{0000-0003-0431-4384}
\affiliation{%
 \institution{Texas A\&M University}
 \country{USA}}
\email{ying.wang@tamu.edu}

\author{Jasper Verheul}
\orcid{0000-0002-2939-8046}
\affiliation{%
 \institution{Cardiff Metropolitan University}
 \country{UK}
}
\email{jpverheul@cardiffmet.ac.uk}

\author{Sang-Hoon Yeo}
\orcid{0000-0002-7140-7954}
\affiliation{%
 \institution{University of Birmingham}
 \country{UK}
}
\email{s.yeo@bham.ac.uk}

\author{Nima Khademi Kalantari}
\orcid{0000-0002-2588-9219}
\affiliation{%
 \institution{Texas A\&M University}
 \country{USA}
}
\email{nimak@tamu.edu}

\author{Shinjiro Sueda}
\orcid{0000-0003-4656-498X}
\affiliation{%
 \institution{Texas A\&M University}
 \country{USA}
}
\email{sueda@tamu.edu}

\renewcommand\shortauthors{Wang, Y. et al.}

\begin{abstract}
We propose a simple and practical approach for incorporating the effects of muscle inertia, which has been ignored by previous musculoskeletal simulators in both graphics and biomechanics.
\editTwo{We approximate the inertia of the muscle by assuming that muscle mass is distributed along the centerline of the muscle.}
We express the motion of the musculotendons in terms of the motion of the skeletal joints using a chain of Jacobians, so that at the top level, only the reduced degrees of freedom of the skeleton are used to completely drive both bones and musculotendons.
Our approach can handle all commonly used musculotendon path types, including those with multiple path points and wrapping surfaces.
For muscle paths involving wrapping surfaces, we use neural networks to model the Jacobians, trained using existing wrapping surface libraries, which allows us to effectively handle the Jacobian discontinuities that occur when musculotendon paths collide with wrapping surfaces.
We demonstrate support for higher-order time integrators, complex joints, inverse dynamics, Hill-type muscle models, and differentiability.
In the limit, as the muscle mass is reduced to zero, our approach gracefully degrades to traditional simulators without support for muscle inertia.
\edit{Finally, it is possible to mix and match inertial and non-inertial musculotendons, depending on the application.}
\end{abstract}

%
%

%
%


\teaser

\maketitle

\section{Introduction}

Computer animation researchers have been using and extending muscle-driven skeletal simulations for many applications---for example, for improved inverse kinematics \cite{Komura2001}, head/neck animation, \cite{Lee2006}, hand animation \cite{Sueda2008}, real-time visualization of muscle activations \cite{Murai2010}, energy-minimizing gait animations \cite{Wang2012}, creation of imaginary bipedal characters, \cite{Geijtenbeek2013}, upper body animations \cite{Lee2009,Si2015}, and control of characters under various anatomical conditions \edit{[\citealt{Lee2014}; \citealt{Lee2019}]}.
However, almost all musculoskeletal simulators used in graphics and biomechanics ignore the effect of the inertia of the muscles as they slide with respect to the bones.
Instead, the mass of the muscles is ``lumped'' to the bones at rest pose, and so the effect of the muscle inertia cannot be reflected in the dynamics of the system, even though around 40\% of total body mass comes from skeletal muscles \cite{Marieb2010}.

\edit{
Missing inertia can change some important aspects of the simulation.
The effect of the missing inertia is most pronounced when the muscle mass is large and far from the joints it acts on.
For example, some of the muscles of the lower limb exhibit significant inertial effects.
In the seminal paper, \citet{Pai2010} notes that the triceps surae muscle of the human ankle can account for an additional $7.6$\% of the effective inertia of the joint.
In \autoref{sec:comparison} (\autoref{fig:teaserRun}), we also show that the combined effect of the muscle mass alters the inverse dynamics result of running motion by as much as 40\%.
As another example, consider the extrinsic muscles of the hand, which are located in the forearm (\autoref{fig:teaserFlick} \& \autoref{sec:stability}).
The joints of the finger have very small inertia by themselves, but when the muscle masses are taken into account, the joint inertia increases significantly.
}
\editTwo{
With a traditional musculoskeletal simulator, these muscle masses are absorbed into the nearest segment (\ie forearm) and do not affect the inertia of the finger joints, whereas with our approach, these masses are coupled to all of the joints spanned by the musculotendons.
}
\edit{
This increase in inertia is important not only for simulation accuracy but also stability.
If we apply an impulse to the fingertip (\eg flicking with the other hand), the distal joint quickly becomes unstable due to its small inertia, but if the effect of muscle inertia is taken into account, it remains stable under an impulse several times larger.
Joint damping can be added to overcome some of these issues, but this would require manual tweaking of parameters, and the added damping would help stabilize both the simulation with and without muscle inertia.
Furthermore, the muscle inertia provides coupling of the joints, naturally preventing the joints from moving independently.
}

\edit{
In the past few years, biomechanics researchers have proposed techniques to deal with muscle inertia \cite{Han2015,Guo2020},%
}
but these approaches can only be used for relatively simple muscle paths.
We therefore propose a framework for incorporating the effects of muscle inertia for more complex muscle path types, including those with wrapping surfaces.
To maximize interoperability with existing musculoskeletal simulators (\eg \cite{Damsgaard2006,Seth2018}), we use the reduced coordinates of the articulated rigid body system representing the skeletal joints as the degrees of freedom.
However, unlike existing musculoskeletal simulators, we take into account the inertia of the muscles as they slide with respect to the bones, by inserting mass points along the paths of the musculotendons.
As the skeleton moves, these mass points move; since each musculotendon is assumed to be frictionless, the path moves such that \edit{its length is minimized}.

Our main technical contribution is the derivation of this mapping (\ie Jacobian, plus its time derivative) from the skeletal motion to the muscle mass motion.
To aid us in the derivation, we categorize musculotendon paths into three types (\autoref{fig:concrete}):
\edit{
\begin{itemize}[topsep=0pt,itemsep=0pt,partopsep=0pt,parsep=0pt]
\item[I:] Straight-line paths, whose Jacobians are derived in a straight-forward manner (\autoref{sec:TypeI}).
\item[II:] Polyline paths through a sequence of points, whose Jacobians are derived by extending the Eulerian-on-Lagrangian framework \cite{Sueda2011,Sachdeva2015} (\autoref{sec:TypeII}).
\item[III:] All others, but most importantly, curved paths wrapping over smooth surfaces, whose Jacobians are based on neural networks trained with our custom sampling strategy to handle parasitic discontinuities (\autoref{sec:TypeIII}).
\end{itemize}
}

To summarize, our contributions are:
\begin{itemize}[topsep=0pt,itemsep=0pt,partopsep=0pt,parsep=0pt]
	\item An Eulerian-on-Lagrangian approach for the inertia of polyline musculotendons composed of a sequence of path points.
	\item A neural network approach for the inertia of curved musculotendons wrapping over smooth surfaces.
	\item A framework compatible with various existing techniques, including higher-order integrators, inverse dynamics, Hill-type muscle models, and differentiability.
	\item A framework capable of handling musculotendons with inertia but can, in the limit, reproduce the results from existing simulators without inertia.
	\item \edit{A framework with support for mixing and matching of inertial and non-inertial muscles, so that the user can choose to add inertia only to muscles with substantial inertial effects.}
\end{itemize}

\figConcrete{}

\section{Related Work}
\label{sec:related}

Because of the importance of human character animation to graphics, many different types of approaches have been studied, starting with the seminal work on facial animation \citep{Waters1987,Terzopoulos1990,Waters1990}.
Often in graphics, the causal relationship between the muscles and the bones is switched---the skeleton is first moved, and then the muscles/flesh are correspondingly simulated to add bulging effects to the character’s skin \cite{Scheepers1997,Wilhelms1997,Kim2011}.
As important as these works are to graphics (\eg commercial products \cite{Maya2011,Ziva2018}), this paper focuses exclusively on muscle-driven systems.

Line-based musculoskeletal methods were developed by adding line-of-action muscles to rigid body dynamics from robotics \cite{Damsgaard2006,Seth2018}.
Almost always, these muscles are assumed to be massless, taking the shortest path between the origin and insertion, possibly being routed around path points and wrapping surfaces.
Perhaps the first work in computer graphics to use proper biomechanics-based muscle models is the work by Komura et al.\ \shortcite{Komura1997,Komura2000,Komura2001}, in which they show new types of animations, such as biomechanically based fatigue, which were not possible with previous joint torque-based approaches.
\citet{Lee2006} use line-based musculotendons to model the muscles of the neck, and in their follow-up works, they use these muscles to drive the volumetric mesh for upper-body motion \cite{Lee2009} and swimming \cite{Si2015}.
\citet{Wang2012} \edit{simulate} a variety of gaits, showing that optimizing for metabolic energy expenditure increases the realism of resulting animations. 
\citet{Geijtenbeek2013} use Hill-type muscle models for a range of bipedal characters, including humans, animals, and imaginary creatures.
Unlike previous work, they also optimize for the placement and routing of these muscle lines so that the total error based on speed, orientation, and effort is minimized.
\citet{Lee2014} propose a scalable biped controller that is able to solve for the activations of more than one hundred muscles.
Their controller is formulated as a quadratic program that can handle frictional contact based on Coulomb's model.
Their results include motions that include muscle pain, muscle tightness, or joint dislocation.
In their follow-up work, \citet{Lee2019} use deep reinforcement learning to control more than three hundred Hill-type muscles for full-body motions.
They show that they can reproduce a wide range of motions, including muscle weakness, use of prostheses, and pathological gaits.


Although not directly related, we briefly cover volume-based muscle models because of their importance to graphics.
Among those that do use biomechanically based muscle mechanics models, two subtypes of volume-based methods have been studied.
The first subtype---those with embedded force generators--- was initially used in animation. 
\citet{Chen1992} introduced the first biomechanics-based muscle mechanics model to computer animation.
They used the finite element method (FEM) with twenty-node isoparametric brick elements, with the longitudinal edges of these elements acting as muscle force generators.
Later, \citet{Zhu1998} used eight-node brick elements with force generators between a set of linear FEM nodes.
\citet{Lemos2001} developed a general FEM framework that could support any nonlinear material as the background isotropic material.
\citet{NgThowHing2001} used a similar approach to embed force generators inside a B-spline solid.
Around the turn of the century, the second subtype---those with anisotropic muscle material models---became more popular in graphics.
The seminal work by \citet{Teran2003} used a material model with a strain energy that includes an anisotropic muscle potential term.
Similar muscle mechanics model is used in their follow-up work on larger scale simulation of skeletal muscles \citep{Teran2005} as well as facial muscles \citep{Sifakis2005}.
\citet{Fan2014} used a blackbox deformation energy as an approximation for contractile mechanics in their volumetric muscles undergoing contact.
Recently, \citet{Lee2018} simulated volumetric muscles with Projective Dynamics, driven by per-element energy functions derived from a Hill-type muscle model.
\editTwo{
\citet{Min2019} used quadratic strain energy to model contractile volumetric muscles of soft-bodied animals.
Although in principle it is possible to use these volumetric simulators to compute the inertial effects of the muscle, they are \textit{impractical or impossible} for the types of applications we are interested in, considering the high number of parameters and the computational complexity required by volumetric models.
}

\section{Methods}
\label{sec:methods}

We use the reduced coordinates, $\qq_r$, of the articulated rigid body system representing the skeletal joints as the degrees of freedom (DOFs) of the system.
To take into account the inertia of the muscles as they slide with respect to the bones, we insert mass points along the path of the musculotendon.
These mass points are \textit{fixed} at a certain percentage length $\alpha$ along the path (\ie fixed at certain texture coordinates; see \autoref{fig:beta}); however, as the skeleton moves, these mass points \textit{move in world space}, since each musculotendon is assumed to be frictionless---the path moves such that its length is \edit{minimized}.

In this section, we derive the Jacobian $\Jar$ that maps the change in the reduced coordinates of the articulated rigid body system to the change in the 3D world coordinates of these muscle mass points:
\begin{equation}
	\label{eq:Jar}
	\dxa = \Jar \dqr,
\end{equation}
where $\dqr$ is the stacked vector of reduced (joint) velocities, and $\dxa$ is the stacked vector of muscle mass point velocities in world space.
The size of $\dqr$ depends on the joint types.
For example, if all of the joints are revolute, then $\dqr \in \rsize{n}$, and if all of the joints are spherical, then $\dqr \in \rsize{3n}$, where $n$ is the number of joints.
The multiplication by the Jacobian $\Jar$, which depends nonlinearly on $\qq_r$, produces the 3D world velocities of muscle mass points $\dxa \in \rsize{3m}$, where $m$ is the number of mass points.

We assume that we already have access to the Jacobian $\Jmr$ (and its time derivative $\dJmr$) that maps between the reduced (joint) velocities and the maximal (body) velocities of the articulated rigid body system \cite{Kim2011,Wang2019}:
\begin{equation}
	\label{eq:Jmr}
	\dqm = \Jmr \dqr,
\end{equation}
where $\dqm$ is the stacked vector of maximal velocities.
Unlike reduced velocities, the size of the maximal velocity vector does not depend on the joint type: $\dqm \in \rsize{6n}$.
In our work, we stack the rotational velocity, $\omega$, and the translational velocity, $\nu$, together to form the maximal velocity, so that for each body, we have:
\begin{equation}
	\dqm = \phi = 
	\begin{pmatrix}
		\omega \\ \nu
	\end{pmatrix},
\end{equation}
with both $\omega$ and $\nu$ expressed in body-local coordinates \cite{Murray2017}.\footnote{Other conventions can be used; the derivations will need to be accordingly modified.}
In the rest of this section, we sometimes use $\phi$ as an alternative symbol for the maximal velocity (twist) of a \textit{single} body.

The main technical contribution of our work is the derivation of Jacobian $\Jam$ (and its time derivative $\dJam$) that maps the maximal velocities to the muscle mass point velocities (details in \autoref{sec:TypeI}, \autoref{sec:TypeII}, and \autoref{sec:TypeIII}).
Once this Jacobian is derived, to compute the world velocities of the muscle mass points from the reduced velocities of the joints, we chain it together with $\Jmr$ to form the final Jacobian we are after:
\begin{equation}
	\Jar = \Jam \Jmr.
\end{equation}
Armed with this Jacobian, we can compute the 3D world accelerations of the muscle mass points as:
\begin{equation}
\begin{split}
	\ddxa &= \dJar \dqr + \Jar \ddqr\\
	\dJar &= \dJam \Jmr + \Jam \dJmr.
\end{split}
\end{equation}
Plugging this into the equations of motion of the mass points $\Ma \ddxa = \fa$ and \edit{applying the principle of virtual work}, we obtain:
\begin{equation}
	\label{eq:muscleFMA}
	\Jar^\top \Ma \Jar \ddqr = \Jar^\top \left( \fa - \Ma \dJar \dqr \right).
\end{equation}
Here, $\Ma \in \Rsize{3m}{3m}$ is the constant diagonal inertia matrix of the $m$ muscle mass points, and $\fa \in \rsize{3m}$ is the force of gravity acting on these mass points.
\edit{The muscle activation forces do not directly apply forces to these mass points.
Instead, in order to keep our framework compatible with existing biomechanical simulators, we assume that the activation forces are applied to the skeleton, which in turn kinematically moves the mass points through the Jacobian $\Jar$.}
The last term in \autoref{eq:muscleFMA}, which uses $\dJam$, is the quadratic velocity vector \edit{(QVV)} that results from the partial derivatives of the kinetic energy \cite{Shabana2013}.

The reduced coordinates of the system also drive the bones, and so combining muscles and bones, we obtain the final equations of motion of the whole \textit{musculoskeletal} system \edit{in reduced coordinates}:%
\begin{subequations}
\begin{align}
	\tilde\MM_r \ddqr &= \tilde\ff_r\\
	\tilde\MM_r &= \Jar^\top \Ma \Jar + \Jmr^\top \Mm \Jmr \label{eq:Mrtilde}\\
	\tilde\ff_r &= \Jar^\top \left( \fa - \Ma \dJar \dqr \right) + \Jmr^\top \left( \fm - \Mm \dJmr \dqr \right) + \ff_r,
\end{align}
\end{subequations}
where $\Mm \in \Rsize{6n}{6n}$ is the constant diagonal inertia of the $n$ bones,\footnote{The maximal inertia is constant because of our choice of body-local coordinates.} $\fm \in \rsize{6n}$ is the sum of maximal forces acting on these bones, such as gravity, Coriolis, and muscle activation forces, and $\ff_r$ is the sum of reduced forces, such as joint torques.
We can use any time integrator to step the system forward in time.
In our implementation, we use forward Euler, BDF1, and SDIRK2 \cite{Hairer2006}.

Throughout this section, we will use the concrete running example shown in \autoref{fig:concrete}.
We will assume that each joint is a revolute joint, and so the reduced velocity is $\dqr = (\dot\theta_A \; \dot\theta_B \; \dot\theta_C)^\top \in \rsize{3}$.
The maximal velocity is $\dqm = (\phi_A \; \phi_B \; \phi_C)^\top \in \rsize{18}$, and $\Jmr \in \Rsize{18}{3}$.
The origin of the musculotendon is assumed to be on body $A$, and the insertion on body $C$.
We will also assume that there is a single muscle with two mass points, so that $\dxa \in \rsize{6}$, and $\Jam \in \Rsize{6}{18}$.
The final Jacobian is $\Jar \in \Rsize{6}{3}$.
For the Type II muscle, the path point is attached to body $B$.
For the Type III muscle, the wrapping surface $S$ is defined with respect to body $B$.

\subsection{Type I: Straight Line Muscles}
\label{sec:TypeI}

We start with the simple case of a straight line muscle between two bodies.
This subsection is not a contribution, but the derivations and notations introduced here will help us with the rest of the paper.

To be explicit, for vectors, we will use a leading superscript to indicate which coordinate space the vector is defined in, and for matrices, we will use a leading sub/superscript to indicate from which to which space the matrix transforms a vector.
Let $^A\xO$ be the 3D position of the origin in the local space of $A$, and $^C\xI$ be the 3D position of the insertion in the local space of $C$.
Then the world velocities of the origin and insertion can be computed as:
\begin{equation}
\label{eq:xOIw}
\begin{split}
	\dxOw = \, \SO{\W}{A} \, \Gamma(^A\xO) \, \phi_A, \quad
	\dxIw = \, \SO{\W}{C} \, \Gamma(^C\xI) \, \phi_C,
\end{split}
\end{equation}
where $\SO{\W}{X} \in SO(3)$ is the rotation matrix of body $X$ (\eg $A$ or $C$), and $\Gamma(\xx) = \begin{pmatrix} [\xx]^\top & \II \end{pmatrix} \in \Rsize{3}{6}$ is the material Jacobian matrix for computing the point velocity \cite{Murray2017}, with $[\cdot]$ the cross-product matrix.
This gives us the following expression for the Jacobian between maximal velocities and world velocities of the origin/insertion \edit{for our concrete running example in \autoref{fig:concrete}}:
\begin{equation}
	\Jxm = 
	\begin{pmatrix}		
	\SO{\W}{A} \, \Gamma(^A\xO) & \Zero & \Zero\\
	\Zero & \Zero & \SO{\W}{C} \, \Gamma(^C\xI)
	\end{pmatrix}
	\in \Rsize{6}{18}.
\end{equation}

For a muscle mass point $\alpha$, the world velocity is simply the weighted average of the world velocities of the origin and the insertion: $\dxaw = (1-\alpha)\,\dxOw + \alpha\,\dxIw$.
Thus, the Jacobian $\Jax$ is:
\begin{equation}
	\Jax =
	\begin{pmatrix}
		(1-\alpha_1) \, \II & \alpha_1 \, \II \\
		(1-\alpha_2) \, \II & \alpha_2 \, \II
	\end{pmatrix} \in \Rsize{6}{6},
\end{equation}
where $\alpha_1$ and $\alpha_2$ are the percentage lengths of the two mass points.
The product of these two Jacobians gives the final Jacobian for Type I muscles: $\Jam = \Jax \Jxm \in \Rsize{6}{18}$.

The $\alpha$ value is fixed over time, as well as the origin and insertion positions with respect to their respective bodies.
The time derivative of the Jacobian is then $\dJam = \Jax \dJxm$, where
\begin{equation}
\label{eq:dJxm}
	\dJxm = 
	\begin{pmatrix}
		\SO{\W}{A} \, [\omega_A] \Gamma(^A\xO) & \Zero & \Zero \\
		\Zero & \Zero & \SO{\W}{C} \, [\omega_C] \Gamma(^C\xI)
	\end{pmatrix},
\end{equation}
since $\dot\RR = \RR [\omega]$ for maximal velocities in body coordinates \cite{Murray2017}.

\subsection{Type II: Path Point Muscles}
\label{sec:TypeII}

\figEOLbeta{}

\figWrapCollision{}

Some musculotendons are constructed as a polyline going through a sequence of path points.
To deal with these types of muscles, we extend the Eulerian-on-Lagrangian (EOL) strands framework \cite{Sueda2011,Sachdeva2015}.
Let $i = 0, 1, 2, \cdots, n+1$ be the indices of the path points (so that $i=0$ corresponds to the origin, $i=n+1$ corresponds to the insertion, and there are $n$ internal path points).
With the EOL framework, we keep track of not only the world space position and velocity (Lagrangian quantities $\xx_i$ and $\dot{\xx}_i \in \rsize{3}$) of the path points, but also the reference space position and velocity (Eulerian quantities $s_i$ and $\dot{s}_i \in \rsize{}$) at these path points.
This allows us to model the sliding motion of the underlying strand even when the world positions of the path points are fixed (\eg if $\dot{\xx}_i = 0$ but $\dot{s}_i \ne 0$, the musculotendon material still moves in world space).
Following the work by \citet{Sachdeva2015}, we assume that all of the line segments of the polyline share the same strain value, which allows us to derive a Jacobian that maps from $\dot{\xx}_i$ to $\dot{s}_i$ (see Eq.~3 \cite{Sachdeva2015}):
\begin{equation}
	\Jsx = -\LL^{-1} \Delta\SS \, \Delta\bar{\XX}, 
\end{equation}
where $\Delta\SS$ is a matrix constructed from the Eulerian coordinates $s_i$, $\Delta\bar{\XX}$ is a matrix constructed from the Lagrangian coordinates $\xx_i$, and $\LL$ is constructed from the segment lengths between the path points.

Since \citet{Sachdeva2015} used \textit{inextensible} EOL strands, they did not need to derive the time derivative of this Jacobian.
However, in this work, the EOL strands are used for \textit{extensible} musculotendons; therefore, we must also derive $\dJsx$.
Using the inverse derivative identity for $\LL$, we obtain:
\begin{equation}
	\dJsx = -\LL^{-1} \left(\dot\LL \, \Jsx + \Delta\dot\SS \, \Delta\bar{\XX} + \Delta\SS \, \Delta \dot{\bar{\XX}} \right).
\end{equation}
Further details are in the supplementary document.

So far, the Jacobians $\Jsx$ and $\dJsx$ that we derived cannot be plugged into our system because they only map between $\dot{\xx}_i$ and $\dot{s}_i$, rather than from $\dqm$ to $\dxa$.
In other words, these Jacobians only provide the mapping between the Lagrangian and Eulerian velocities of the path points of a musculotendon, rather than the mapping between the maximal velocities of the skeleton and the muscle mass point velocities.
To tie the Jacobians $\Jsx$ and $\dJsx$ to the rest of the system, we introduce a new notation $\zz$ that represents the combined Lagrangian/Eulerian coordinates:
\begin{equation}
	\zz_i =
	\begin{pmatrix}
		\xx_i \\
		s_i
	\end{pmatrix} \in \rsize{4}.
\end{equation}
In the concrete example in \autoref{fig:concrete}, which contains a single internal path point, $\zz = (\xO \; \sO \; \xx_1 \; s_1 \; \xI \; \sI)^\top \in \rsize{12}$.
The musculotendon material cannot flow past the origin or insertion, so $\dot{s}_\text{ori}$ and $\dot{s}_\text{ins}$ are always zero.
Using this notation, the Jacobian that we are after can be written as:
\begin{equation}
\label{eq:JazJzm}
\begin{split}	
	\Jam &= \Jaz \Jzm \in \Rsize{6}{18}\\
	\dJam &= \dJaz \Jzm + \Jaz \dJzm.\\
\end{split}
\end{equation}

The left Jacobian $\Jaz \in \Rsize{6}{12}$ represents the mapping from the Lagrangian/Eulerian velocities of the path points to the muscle mass point (\autoref{fig:EOL}).
This was already derived by \citet{Sueda2011} (Eq.~4), but we reproduce the expression here, for our concrete example with one path point and two mass points.
The first mass point is between the origin and the path point, and the second mass point is between the path point and the insertion.
Therefore, we get:
\begin{equation}
\setlength{\arraycolsep}{4pt}
\begin{split}
	&\Jaz = \\
	&\begin{pmatrix}
		(1-\beta_1) \II & -(1 - \beta_1) \FF_1 & \beta_1 \II & -\beta_1 \FF_1 & \Zero & \Zero\\
		\Zero & \Zero & (1-\beta_2) \II & -(1 - \beta_2) \FF_2 & \beta_2 \II & -\beta_2 \FF_2
	\end{pmatrix}.
\end{split}
\end{equation}
Here, we used $\beta$ to represent the percentage location of $\xx_\alpha$ \textit{within} a particular line segment, as shown in \autoref{fig:beta}.
$\FF \in \rsize{3}$ is the deformation gradient of the line segment: $\FF_1 = (\xx_1 - \xO) / (s_1 - \sO)$ and $\FF_2 = (\xI - \xx_1) / (\sI - s_1)$.
The time derivatives of these quantities, which were not derived before by \citet{Sueda2011}, are nevertheless needed for our extensible musculotendons.
We list the detailed derivations of these derivatives in the supplementary document.

The right Jacobian $\Jzm \in \Rsize{12}{18}$ in \autoref{eq:JazJzm} represents the mapping from the maximal velocities of the bodies to the Lagrangian/Eulerian velocities of the path points.
This can be accomplished by constructing a Jacobian that passes through the Lagrangian components while hitting the Eulerian components by $\Jsx$:
\begin{equation}
	\Jzm =
	\begin{pmatrix}
		\II\\
		\Jsx
	\end{pmatrix}
	\Jxm, \quad
	\dJzm =
	\begin{pmatrix}
		\Zero\\
		\dJsx
	\end{pmatrix}
	\Jxm + 
	\begin{pmatrix}
		\II\\
		\Jsx
	\end{pmatrix}
	\dJxm.
\end{equation}
$\Jxm$ in our concrete example with an internal path point $\xx_i$ attached to body $B$ is:
\begin{equation}
\renewcommand\arraystretch{1.5}
	\Jxm = 
	\begin{pmatrix}		
	\SO{\W}{A} \, \Gamma(^A\xO) & \SO{\W}{B} \, \Gamma(^B\xx_i) & \Zero\\
	\Zero & \SO{\W}{B} \, \Gamma(^B\xx_i) & \SO{\W}{C} \, \Gamma(^C\xI)
	\end{pmatrix}
	\in \Rsize{6}{18}.
\end{equation}
Its time derivative, $\dJxm$, can be derived similarly as in \autoref{eq:dJxm}.

\subsection{Type III: Wrapping Surface Muscles}
\label{sec:TypeIII}

\edit{
Some musculotendons are constructed as 3D paths that wrap around smooth surfaces.
To derive the Jacobians for these types of paths, we use neural networks.}
The reason for using neural networks may not be immediately obvious, since existing muscle routing algorithms are highly efficient \cite{Garner2000,Scholz2016,Seth2018,Lloyd2020}.
With some fairly minor modifications, we could use the output of these libraries to compute the Jacobians with finite differencing, which would not be prohibitively expensive due to the efficiency of these libraries.
However, they cannot be used directly in our framework for inertial muscles because they all suffer from a massive problem: \textit{Jacobian discontinuity}.

As an illustration of this problem, suppose that we have a double pendulum with a musculotendon shown in \autoref{fig:wrapAnim}.
As the pendulum swings due to the force of gravity acting on both the bones and the musculotendon, the path of the musculotendon attaches and detaches from the wrapping surface.
If we use a Jacobian computed using existing wrapping surface libraries and finite differencing, we observe discontinuities in the energy plot, as shown in \autoref{fig:wrapEnergyExisting}.
\edit{These energy jumps occur because the velocities of the muscle mass points undergo sudden changes, even when the velocities of the joints vary smoothly.}
\autoref{fig:wrapX} shows the x-component of five of the mass points (each with its own color), as a function of the distal joint angle, zoomed in near a discontinuity.
The values computed with an existing wrapping surface library are shown with solid lines, and ours with dotted lines.
\autoref{fig:wrapJ} shows the corresponding derivatives.
The jump in the value of the Jacobian \edit{creates sudden changes in the velocities of the mass points, which in turn creates} energy jumps in the simulation.
On the other hand, our neural network approach generates the smooth Jacobian plots in \autoref{fig:wrapJ}, while keeping the position plots in \autoref{fig:wrapX} virtually indistinguishable from the output of the library code.
This results in a smooth energy trajectory shown in \autoref{fig:wrapEnergyOurs}.

One way to deal with the discontinuity is to detect these sudden state changes and apply a manual fix, \eg by computing the pre- and post-collision Jacobians and running a nonlinear optimization to compute the velocities that minimize the change in energy.
However, such approaches are tricky to incorporate into implicit integrators, such as SDIRK2 \cite{Hairer2006}, as well as into differentiable simulation techniques, such as the adjoint method \cite{McNamara2004,Geilinger2020,Xu2021}, \edit{which our method supports naturally without any changes to the framework}.

We instead choose to smooth the discontinuity.
Smoothing would be easy with a uni-articular muscle spanning a hinge joint.
As an offline process, we could pre-sample many points within the range of motion of the joint, and then apply a smoothing filter over the samples.
During runtime, we could then use the filtered values to construct the Jacobian.
However, high-dimensional smoothing would be required with a bi- or multi-articular muscle, as well as with a uni-articular muscle with a spherical joint.
Therefore, we use neural networks for this high-dimensional smoothing problem.
This approach is simple to implement and can be used with any existing muscle routing libraries.

\subsubsection{Training the Network}

We train the network with origin and insertion positions as the input, rather than the joint angle.
This is an important choice, since it allows the same trained network to be used regardless of the type of the joints, how many joints the musculotendon spans, as well as with respect to which bodies the surface is defined.
Using the cylinder wrapping surface as a concrete example, the input and output of our network are:
\begin{equation}
	\begin{pmatrix}
		^S\xO\\
		^S\xI\\
		\alpha\\
		r
	\end{pmatrix}
	\rightarrow
	\begin{pmatrix}
		^S\xx_\alpha
	\end{pmatrix},
\end{equation}
where $r$ is the radius of the cylinder, and $\alpha$ is the percentage length along the musculotendon.
The origin $^S\xO$, insertion $^S\xI$, and the output position $^S\xx_\alpha$ are all defined with respect to the coordinate space of the wrapping surface $S$.
During training, we use the $\ell^2$-norm of the difference between the output of the network and the output of the wrapping library.
\edit{We include samples with muscles in both attached and detached states, so that at runtime, we do not need to detect whether the muscle is in contact or not.}
Once trained, the network and the original wrapping library can be used interchangeably, except for one important difference: discontinuity.

To ensure that the network does not contain any discontinuities, we use the \textit{hyperbolic tangent} activation function.
Furthermore, we \textit{throw away} the samples near the discontinuity before training.
To detect whether a sample is close to a discontinuity, we use the following simple heuristics for all wrapping surfaces.
\begin{itemize}[topsep=0pt,itemsep=0pt,partopsep=0pt,parsep=0pt,leftmargin=15pt]
	\item Compute $l$, the length of the ``wrapped'' portion of the path.
	\item If $l = 0$, keep the sample.
	\item Compute $L$, the length of the whole path.
	\item If $l/L < \text{thresh}$, discard the sample.
	\item Otherwise, keep the sample.
\end{itemize}
Both $l$ and $L$ are readily available from the wrapping surface library.
In our current implementation, we use a threshold of 1\%.

The trajectory of $\xx_\alpha$ computed with the library is only $C^0$, but the trajectory computed by the network is $C^\infty$.
Despite this difference, the two trajectories are virtually indistinguishable.
For example, if we closely inspect what happens to $\xx_\alpha$ as it approaches and touches the wrapping surface, we find that it slightly penetrates the surface and then floats back to the surface.
We also note that the wrapping surface path is already an approximation of the actual path taken by a real muscle, and so this slight discrepancy is within reason.

\subsubsection{Incorporating the Network}

We now describe how we use the trained network in our simulation framework.
As described earlier, to maximize generality, we train the network with origin and insertion in the coordinate space of the wrapping surface as the input: $\xOs$ and $\xIs$.
To compute the world velocity of the muscle mass point, $\dxaw$, we first need to transform the network input into $S$ space, use the network, and then transform the output back to world space.

Like with Type I and Type II muscles, our goal is to derive $\Jam$ and $\dJam$.
To derive $\Jam$, we must express the world velocity of $\xx_\alpha$ using maximal velocities of the bodies.
The world velocity of one mass point can be written as the sum of three terms:
\begin{equation}
\label{eq:dxawTerms}
	\dxaw = \, \vBw + \, \edit{\vOw} + \, \edit{\vIw}.
\end{equation}
The first term represents the \textit{base motion} of the mass point as if it were fixed with respect to $S$. 
Since $S$ itself could be moving, even if the mass point is stationary in $S$, its world velocity could be nonzero.
The second term represents the contribution from the relative motion of the \textit{origin} within the $S$ space.
Similarly, the third term represents the contribution from the relative motion of the \textit{insertion} within the $S$ space.
Our goal is to rewrite each of the three terms so that $\dxaw = \Jbase \dqm + \Jori \dqm + \Jins \dqm$.
Then the Jacobian we are after is $\Jam = \Jbase + \Jori + \Jins$.

For concreteness, we continue to assume that the origin is fixed to $A$, insertion is fixed to $C$, and the surface $S$ is fixed to $B$ (see \autoref{fig:concrete}).
The first term in \autoref{eq:dxawTerms} is the motion of the mass point assuming that it is fixed in $S$.
If we convert this to body $B$'s space, we get:
\begin{equation}
\begin{split}	
	\vBw &= \SO{\W}{S} \, \Gamma(\xas{}) \, \phi_S \\
	&= \SO{\W}{B} \, \Gamma(\SE{B}{S} \, \xas{}) \, \phi_B,
\end{split}
\end{equation}
where $\SE{B}{S}$ is the transformation matrix of $S$ with respect to $B$, which is fixed over time.
The Jacobian for this term, assuming there are two mass points (\autoref{fig:concrete}), is then
\begin{equation}
	\renewcommand\arraystretch{1.5}
	\Jbase =
	\begin{pmatrix}
		\Zero & \SO{\W}{B} \, \Gamma(\SE{B}{S} \, \xas{1}) & \Zero\\
		\Zero & \SO{\W}{B} \, \Gamma(\SE{B}{S} \, \xas{2}) & \Zero
	\end{pmatrix} \in \Rsize{6}{18},
\end{equation}
where $\xas{1}$ and $\xas{2}$ are the \textit{values returned from the network}.

\figWrapSpaces{}

To compute $\Jori$, we first need the relative velocity of the origin from the point of view of the surface.
To do so, we must take into account the relative motions of the coordinate spaces, shown in \autoref{fig:wrapSpaces}.
Since the origin is attached to $A$, we can compute its world velocity $\dxOw$ using \autoref{eq:xOIw}.
What we are after is the relative velocity of the origin if we temporarily imagine frame $B$ to be stationary and transfer its motion to frame $A$.
In other words, we subtract from $\dxOw$ the hypothetical velocity of the origin attached to body $B$:
\begin{equation}
\label{eq:vOwrel}
	\vOwrel = \SO{\W}{A} \, \Gamma(^A\xO) \, \phi_A - \SO{\W}{B} \, \Gamma(\SE{B}{A} \, ^A\xO) \, \phi_B,
\end{equation}
where $\SE{B}{A} = \SE{\W}{B}^{-1} \SE{\W}{A}$, formed from the current configurations of bodies $A$ and $B$.
We then rotate this into surface space, hit it with the \textit{network Jacobian}, and then rotate back to world:
\begin{equation}
\label{eq:vOw}
	\vOw = \SO{\W}{S} \, \JaO{} \, \SO{S}{\W} \, \vOwrel.
\end{equation}
The network Jacobian, $\JaO{}$, is computed with backward differentiation of the network.
Given that the input and output of the network are in $S$ space, the network Jacobians are also in $S$ space.
\begin{equation}
	\JaO{} = \frac{d \, \xas{}}{d \, \xOs}, \quad \JaI{} = \frac{d \, \xas{}}{d \, \xIs}.
\end{equation}
Since $\xas{}$, $\xOs$, and $\xIs$ are all in $\rsize{3}$, these network Jacobians are $3 \times 3$ matrices.

Combining \autoref{eq:vOwrel} and \autoref{eq:vOw} and extracting out the maximal velocities $\phi_A$ and $\phi_B$, the Jacobian $\Jori$ for the concrete running example becomes:
\begin{equation}
\renewcommand\arraystretch{1.5}
\begin{split}
	\Jori &= 
	\begin{pmatrix}
		\SO{\W}{S} \, \JaO{1} \, \SO{S}{\W} \, \SO{\W}{A} \, \Gamma(^A\xO) & \Zero & \Zero\\
		\SO{\W}{S} \, \JaO{2} \, \SO{S}{\W} \, \SO{\W}{A} \, \Gamma(^A\xO) & \Zero & \Zero
	\end{pmatrix}\\
	&-
	\begin{pmatrix}
		\Zero & \SO{\W}{S} \, \JaO{1} \, \SO{S}{\W} \, \SO{\W}{B} \, \Gamma(\SE{B}{A} \, ^A\xO) & \Zero\\
		\Zero & \SO{\W}{S} \, \JaO{2} \, \SO{S}{\W} \, \SO{\W}{B} \, \Gamma(\SE{B}{A} \, ^A\xO) & \Zero
	\end{pmatrix}
	\in \Rsize{6}{18}.
\end{split}
\end{equation}
The Jacobian $\Jins$ is derived similarly, except that the insertion is fixed to body $C$ instead of $A$.
\begin{equation}
\renewcommand\arraystretch{1.5}
\begin{split}
	\Jins &= 
	\begin{pmatrix}
		\Zero & \Zero & \SO{\W}{S} \, \JaI{1} \, \SO{S}{\W} \, \SO{\W}{C} \, \Gamma(^C\xI) \\
		\Zero & \Zero & \SO{\W}{S} \, \JaI{2} \, \SO{S}{\W} \, \SO{\W}{C} \, \Gamma(^C\xI)
	\end{pmatrix}\\
	&-
	\begin{pmatrix}
		\Zero & \SO{\W}{S} \, \JaI{1} \, \SO{S}{\W} \, \SO{\W}{B} \, \Gamma(\SE{B}{C} \, ^C\xI) & \Zero\\
		\Zero & \SO{\W}{S} \, \JaI{2} \, \SO{S}{\W} \, \SO{\W}{B} \, \Gamma(\SE{B}{C} \, ^C\xI) & \Zero
	\end{pmatrix}
	\in \Rsize{6}{18}.
\end{split}
\end{equation}

The time derivatives of the individual quantities in $\dJbase$, $\dJori$, and $\dJins$ are listed in the supplementary material.
We analytically derive all of the derivatives, except for the network Jacobians.
For these, we perturb $\xOs$ and $\xIs$ in time to evaluate the network again to perform finite differencing:
\begin{equation}
\label{eq:xsFD}
\begin{split}
	\xOs^+ &= \, \xOs + \epsilon \, \vOsrel, \quad
	\dJaO{} = \left( ^S\JJ_{{\alpha}\text{o}}^\text{NN+} - \, \JaO{} \right) / \epsilon,\\
	\xIs^+ &= \, \xIs + \epsilon \, \vIsrel, \quad
	\dJaI{} = \left( ^S\JJ_{{\alpha}\text{i}}^\text{NN+} - \, \JaI{} \right) / \epsilon,
\end{split}
\end{equation}
where $\vOsrel$ is computed as $\vOsrel = \SO{S}{\W} \, \vOwrel$, and likewise for $\vIsrel$.

\section{Results}
\label{sec:results}

We implemented a prototype in MATLAB.
The networks were trained on a computer with a \editTwo{Ryzen 7 5800X} CPU with \editTwo{32} GB of RAM and \editTwo{an RTX 3080 Ti GPU with 12 GB of RAM}.
We trained the networks using Adam~\cite{Kingma14} with the default parameters and a learning rate of $10^{-4}$.
For each network, we used \editTwo{6} layers with 256 neurons per layer.
We used $\tanh$ as the activation function for all layers.
The trained networks were loaded and evaluated in MATLAB.
We used around \editTwo{30k} samples, and the training took about \editTwo{12} hours.
\editTwo{More details are in the supplementary document.}

\subsection{Comparison to Analytical Results}

\figDinesh{}

We start with comparisons to the simulation and analytical results by \citet{Pai2010} to verify that our general framework is in agreement with published results.
First we simulate the scene in \autoref{fig:dineshSim}, which uses the same setup as \textit{their} Fig.~2.
As shown by the solid lines in \textit{our} \autoref{fig:dineshPlot}, the two angles reach zero at 0.3 seconds, just like in the published result.

\citeauthor{Pai2010} also analytically computed the contributions to the self-inertia of the rat knee joint from the biceps femoris posterior muscle and the bones of the shank, and reported that the relative contribution from the muscle with respect to the bones is 45\%.
We also computed the inertia from the muscle and the bones using \autoref{eq:Mrtilde}, and obtained the value of 45.8\%.
The slight discrepancy goes down if we include more mass points, but we found that 10-20 are sufficient for most purposes.
Furthermore, the discrete approach allows us to more easily model the non-uniform mass distribution along the musculotendon path.

\subsection{Network Jacobian}

\figCylinders{}

We simulate a group of double pendulums with varying origin, insertion, radius, and the initial rotation of the wrapping surface, as shown in \autoref{fig:cylinders}.
In these experiments, the masses of the proximal bone, the distal bone, and the muscle are set to be equal.
For comparison, in the right-most pendulum, we remove the muscle, adding half of its mass to the proximal bone and the other half to the distal bone.
Using the same trained network, the simulator is able to account for all the variations properly.

\subsection{Energy Behavior}

\figQvv{}
\figRun{}
\figTorques{}

To show the importance of the $\dJam$ term that we derived, we take one of the simulations from \autoref{fig:cylinders}, and remove $\dJam$, and consequently, the quadratic velocity vector (QVV) of the muscle mass points \cite{Shabana2013}.
(We keep the QVV of the bones in the simulation.)
As shown in \autoref{fig:QvvOffTypeIII}, even with the SDIRK2 time integrator, the energy oscillates wildly.
On the other hand, as shown in \autoref{fig:QvvOnTypeIII}, the energy stays stable once we put the QVV of the muscle back in.
Similarly, in Fig.~\ref{fig:QvvOffTypeII}-\ref{fig:QvvOnTypeII}, we show the same experiment with a Type II muscle.
Again, without the QVV of the muscle, the energy fluctuates, but with the QVV of the muscle included, the energy remains stable.

\subsection{\edit{Simulation Stability}}
\label{sec:stability}

\edit{
The effect of muscle inertia is stronger when a relatively light bone is actuated by a relatively large muscle mass located away from the joint.
In \autoref{fig:teaserFlick}, we show an example of such a case with the flexor digitorum profundus and superficialias muscles (FDP \& FDS), which originate near the elbow and insert into the distal and middle phalanges, respectively.
For our simulation, we modeled the bones and joints using open source data \cite{Lee2015}, and we manually modeled the FDP and FDS as Type II muscles, with the tendons routed through pulleys implemented as path points.
We fixed all joints except for the three joints of the index finger, which we modeled as revolute joints.
The masses of the bones are set from the meshes, with a relatively large density of 5~\si{g.cm^{-3}} to account for the rest of the finger mass, and the mass of the muscles is set to 200~\si{g} each.
With a fixed time step of 1~\si{ms}, we apply different amounts of force for the first two time steps of the simulation, to model flicking the fingertip with the other hand.
\editTwo{
With the traditional approach, the simulation becomes unstable when the force is increased to 5~\si{N}, whereas with our approach, the simulation becomes unstable when the force is increased to 20~\si{N}.
This is due to the fact that with the traditional approach, the muscle inertia gets absorbed into the forearm segment, and thus the generalized inertia of the finger joints is not affected by the muscles, unlike with our approach.
}
(The peak force during typing is around 2~\si{N} \cite{Kim2014}.)
We also note that the inertia due to the muscles in this particular example is \textit{substantially underestimated}, since we assume that strain is equal throughout the length of the musculotendon.
If we also take into account the fact that the tendon is highly stiff, joint motion would cause more of the muscle mass to move, which would increase the inertia further.
}

\subsection{Comparison to OpenSim}
\label{sec:comparison}

For our next experiment, we use \editTwo{marker-based} motion-capture data to drive the skeleton and compute the resulting torques at the joints with inverse dynamics.
We show a 0.5 second clip in \autoref{fig:run}.
The figure shows the swing phase: from \editTwo{take-off} to touch-down.
We use OpenSim to \editTwo{scale} the bone lengths/masses, joint locations, muscle origin/insertion, path points, and wrapping surfaces to the specific subject.
The skeleton has 11 DOFs: 6 for pelvis, 3 for the right hip, 1 for the right knee, and 1 for the right ankle.
We model four muscles that span the ankle: gastrocnemius lateral (Type III), gastrocnemius medial (Type III), soleus (Type I), and tibialis anterior (Type II).
The subject runs on a treadmill at \editTwo{19.1~\si{km/h}}, and we use OpenSim to reconstruct the motion of the skeleton from the marker data.
We collect the ankle torque computed with inverse dynamics from the swing phases from two 10-second trials using OpenSim and our simulator.
\edit{We overlay the swing phases on top of each other and plot the results in \autoref{fig:torques}.}
We show the torque results generated by:
\begin{itemize}[topsep=0pt,itemsep=0pt,partopsep=0pt,parsep=0pt,leftmargin=15pt]
	\item OpenSim ({\color{MATBLUE}blue}), which does not support inertial muscles.
	\item Our simulator ({\color{MATRED}red}) with the muscles accounting for 0\% of the total mass of the tibia segment.
	\item Our simulator ({\color{MATYELLOW}yellow}) with 80\% of the tibia mass transferred to the muscles.
\end{itemize}
The relative masses of the four muscles are taken from the literature \cite{Ward2009}.
For each muscle, the mass is distributed into 20 equally spaced points in the middle portion of the musculotendon that correspond to the muscle (as opposed to the tendons).
\autoref{fig:torques19.1zoom} shows the closeup of the final dip.
Comparing the {\color{MATBLUE}blue} and {\color{MATRED}red} plots, we confirm that our simulator generates results that \textit{gracefully degrades} to OpenSim's results, as the inertia of the muscles are decreased to zero.
On the other hand, comparing the {\color{MATRED}red} and {\color{MATYELLOW}yellow} plots, we note that the ankle moment can differ by as much as 40\% due to the effect of muscle inertia.
\editTwo{In the supplementary material, we show how our result gracefully degrade to OpenSim’s result.}

\subsection{Spline Joint Knee with Hill-Type Muscles}
\label{sec:splineHill}

\figKnee{}
\figKneeAnim{}

To demonstrate the generality and flexibility of our approach, we take the same scene setup as above, but replace the revolute joint of the knee with a spline joint \cite{Lee2008} and add the semimembranosus (Type I) and the rectus femoris (Type III).
As shown in \autoref{fig:knee}, we manually model a spline joint to better model the motion of the tibia with respect to the femur.
(OpenSim uses a similar technique called a ``mobilizer'' \cite{Seth2010}.)
We also use Hill-type muscles \cite{Zajac1989} to drive the knee and ankle joints, \edit{as opposed to using mocap as in \autoref{sec:comparison}.}
We use the damped equilibrium model with active force-length, active force-velocity, passive force-length, and tendon force-length curves taken from the biomechanics literature \cite{Millard2013}.
We manually set the excitation levels of the gastrocnemius lateral/medial muscles to a low level.
We use a proportional controller based on the ankle joint angle to set the excitation levels of the tibialis anterior and the soleus muscles.
Then we manually excite the rectus femoris and semimembranosus muscles, which results in the extension and flexion of the knee, as shown in \autoref{fig:kneeAnim}.

\subsection{Differentiable Reaching with Adjoint Method}

For the final result, we use the adjoint method \cite{McNamara2004,Geilinger2020,Xu2021} to compute the simulation derivatives to optimize for a reaching task using an arm model \cite{Chadwick2014} with manually placed muscles, shown in \autoref{fig:teaserReach}.
For the three heads of the deltoid muscle, we use sphere-capped cylinders, and for the three heads of the triceps brachii muscle, we use cylinders.
The task objective is to move the hand to the specified target, and the task parameters are the constant torques to be applied to the shoulder (3 DOF) and elbow (1 DOF) joints.
We use \texttt{fminunc} as the optimizer with our analytical derivatives.
As a comparison, when we run \texttt{fminunc} in gradient-free mode, it takes an order-of-magnitude more time to optimize, requiring many more simulation runs.
Our inertial muscles, however, work seamlessly with the adjoint method.
Furthermore, more objectives can be added, such as having the hand come to a rest, or more generally, following a preset trajectory.

\section{Conclusion \& Future Work}
\label{sec:conclusion}

We presented an approach to account for the inertia of the muscles in a musculoskeletal simulation.
We are able to handle a wide variety of musculotendon paths, including (I) straight, (II) polyline, and (III) curved paths over wrapping surfaces.
For Type II muscles, we use the Eulerian-on-Lagrangian framework, and for Type III muscles, we use neural networks.
Our approach is compatible with existing simulation techniques, such as inverse dynamics and differentiable dynamics, \edit{and the motion can be driven by muscle activations or joint torques.}
In the limit, as the mass of the muscles is transferred to the bones, our simulation results gracefully degrade to results obtained using traditional musculoskeletal \edit{simulators} without inertial muscles.
\edit{Finally, it is possible to mix and match inertial and non-inertial musculotendons, depending on the application.}

\editTwo{
We use the centerline to account for the muscle mass, which is still an approximation, but this is a prudent choice, since using a full, volumetric mesh is impractical for these experiments, at least currently.
For example, it would be a challenge to produce results with FEM that can gracefully degrade to OpenSim results the way our method can.
It may be possible to tweak the FEM simulation parameters to produce the desired output, but we believe that using FEM for these target applications is extremely challenging if not impossible, considering the high number of parameters and the computational complexity required by the volume model.
Future work may address these difficulties with volumetric FEM.
We believe that such work, along with ours, would pave the way toward a fully comprehensive simulation framework.
}

Some models use path points that move based on the skeletal DOFs (\eg LBS waypoints \cite{Ryu2021}, moving muscle points \cite{Seth2018}).
Although we have not implemented these, they can be categorized as Type II path points with their corresponding Jacobians between the skeletal DOFs and these points.

\edit{
For muscles with long tendons, our approach still underestimates the muscle inertia because we assume that the strain is equal along the entire length of the musculotendon.
For future work, we would like to derive the kinematics of the muscle points while incorporating inextensible tendons to reduce this underestimation.
}

We plan to train on more wrapping surface types, including ellipsoid, torus, sphere, and double cylinder \cite{Seth2018,Garner2000}.
In theory, our neural network approach can be used for any path.
However, some wrapping surfaces require many parameters, which could make training more difficult and slower.
For example, to train a double cylinder, it would require five more parameters than a single cylinder.
(The first cylinder can be defined along the Z-axis.
Assuming that the second cylinder is not orthogonal to the first, we need two parameters for a point and two for the direction, plus the radius.)
Similarly, using a network for an arbitrary shape \cite{Lloyd2020} could be a challenge, depending on the number of parameters of the surface.

Network evaluation is a bottleneck in our current implementation, which is written in MATLAB.
We expect that evaluating the network on the GPU and batching the input as much as possible would increase the performance significantly.
Furthermore, since our framework allows mixing and matching of inertial and non-inertial musculotendons (\eg \autoref{sec:comparison}), it is possible to find a subset of musculotendons to add inertia to, in order to find the sweet spot in terms of efficiency and efficacy.
Automatically determining the set of musculotendons that affects the total inertia the most is an interesting avenue of future research.

Finally, given that our approach is compatible with the adjoint method, it would be interesting to optimize for tasks involving ground contact \cite{Geilinger2020,Xu2021}.
In our current implementation, as with most other musculoskeletal simulators \cite{Millard2013}, musculoskeletal dynamics and \edit{muscle/tendon dynamics} are integrated separately, and so the adjoint method cannot use muscle excitations as parameters.
Going further, we could add another layer on top of the adjoint method to compute for the muscle excitations rather than joint torques.

\begin{acks}

We thank the anonymous reviewers for their helpful comments.
This work was sponsored in part by the National Science Foundation (CAREER-1846368) and by Biotechnology and Biological Sciences Research Council (BB/S003762/1).

\end{acks}

\bibliographystyle{ACM-Reference-Format}
\bibliography{muscle-jacobians}

\end{document}